\documentclass[9pt,twocolumn,twoside]{opticajnl}
\journal{opticajournal} 

\setboolean{shortarticle}{false}

\usepackage{lineno}
\usepackage{amsmath,amssymb,commath,siunitx} 
\newcommand{\SIadj}[2]{\SI[number-unit-product={\text{-}}]{#1}{#2}}
\newcommand{\SIrangeadj}[3]{\numrange[range-phrase=--]{#1}{#2}-\si{#3}}
\DeclareSIUnit\bar{bar}

\usepackage[version=4]{mhchem} 

\newcommand{\etal}[1]{\mbox{{#1} \textit{et al.}}}
\newcommand{\ie}{\textit{i}.\textit{e}.,\ }
\newcommand{\eg}{\textit{e}.\textit{g}.,\ }

\newcommand*\diff{\mathop{}\!\mathrm{d}}

\title{Partial parabolic amplification in rare-earth-doped optical fiber}

\author[1,$\dagger$]{Wenchao Wang}
\author[1,$\dagger$,*]{Yi-Hao Chen}
\author[1]{Frank Wise}
\affil[1]{School of Applied and Engineering Physics, Cornell University, Ithaca, New York 14853, USA}
\affil[$\dagger$]{These authors contributed equally to this work.}

\affil[*]{Corresponding author: yc2368@cornell.edu}

\begin{abstract}
Nonlinear amplification is a powerful technique for generating ultrashort laser pulses with high peak power in fiber systems. However, the diversity of nonlinear amplification approaches and their inherent complexities present significant challenges to achieving a unified understanding and further scaling of peak power and pulse energy while preserving ultrashort durations. Here, we report the results of a systematic optimization with respect to seed pulse duration that elucidates the dynamics of nonlinear amplification and allows identification of distinct propagation regimes. As part of this analysis, we identify a new regime, termed partial parabolic amplification, which achieves \SIadj{50}{\fs} pulse duration and yields higher peak power than any other nonlinear amplification regime known to date. An initial experimental demonstration of partial parabolic amplification produces \SIadj{50}{\fs} and \SIadj{2.2}{\micro\joule} pulses with a \SIadj{25}{\micro\m}-core \ce{Yb} fiber amplifier, corresponding to a \SIadj{30}{\MW} peak power. In contrast to other nonlinear amplification techniques, practical energy scaling beyond \SI{10}{\micro\joule} and \SI{200}{\MW} should be achievable with available gain fibers with larger mode areas, which would fill a gap in existing fiber laser capabilities that would directly impact material processing, nonlinear bio-imaging, and other applications.
\end{abstract}

\setboolean{displaycopyright}{false} 
\begin{document}

\maketitle

\section{Introduction}
With the growing adoption of ultrafast lasers in science and industry, there is a growing demand for energetic pulses with sub-\SIadj{100}{\fs} durations. Such high-peak-power sources are critical for a range of advanced applications, such as high-precision micro-machining \cite{Yoshino2008}, multiphoton microscopy \cite{Horton2013}, and nonlinear spectroscopy\cite{Picque2019}. Prevailing approaches for the generation of high-peak-power ultrashort pulses involve chirped-pulse amplification (CPA) \cite{Strickland1985,Eidam2010,Zhao2015}, multipass \cite{Lowdermilk1980} and regenerative amplification \cite{Raybaut2003,Ueffing2016}, as well as a combination of these approaches. However, gain narrowing in these systems imposes limitations on the amplifiable spectral bandwidth, thereby restricting the achievable pulse duration to several hundred femtoseconds \cite{Raybaut2005,Kuznetsova2007}. Specialized techniques such as use of a gain-flattening filter \cite{Stark2021} or an adaptively-controlled fiber Bragg grating stretcher \cite{Lampen2023} facilitate pulse generation with around \SIadj{100}{\fs} duration. Nonlinear post-compression through self-phase modulation (SPM) or soliton compression can compress the amplified pulse to few cycles \cite{Mak2013}. These techniques overcome gain narrowing but add complexity and extra loss to the system.

A number of techniques exploit the Kerr nonlinearity during the amplification process to simultaneously achieve high pulse energy and short duration. In nonlinear CPA \cite{Zhou2005} and cubicon \cite{Shah2005} systems, accumulated nonlinear phase can compensate for third-order dispersion (TOD) introduced by use of mismatched stretcher and compressor. Nonlinear spectral broadening of a short pulse can be used to counter gain-narrowing \cite{Pouysegur2014,Pouysegur2015}. However, these two techniques generate a pulse whose duration is only comparable to its seed pulse. Self-similar amplification (SSA) produces significantly-shorter pulses than the injected seed through nonlinear evolution to a parabolic shape, where pulse duration, bandwidth, and energy are all constrained to increase in concert with one another \cite{Anderson1993,Tamura1996,Fermann2000,Kruglov2002}, but achieves only \SIadj{\sim200}{\nano\joule} energy with \SIadj{80}{\fs} duration due to the finite gain bandwidth \cite{Limpert2002}. Parabolic pre-shaping decouples the pulse parameters from the self-similar evolution while transforming nonlinear phase accumulation into a compressible linear chirp \cite{Schreiber2006,Pierrot2013,Fu2017}. Pulses as short as \SI{275}{\fs}, compressed from a \SIadj{9}{\ps} pulse, have been demonstrated at \SIadj{4.3}{\micro\joule} energy; however, this approach remains insufficient for reaching sub-\SIadj{100}{\fs} durations. So-called pre-chirped amplification was introduced \cite{Chen2012,Zhao2014,Liu2016b,Song2017,Luo2018,Chang2019,Zhang2020a,Zhang2021b} and produces pulses as short as \SI{24}{\fs} \cite{Song2017}, but it requires an additional stage of precise chirp control of the seed pulse. Recently, gain-managed nonlinear amplification (GMNA) emerged as a simple technique to generate sub-\SIadj{40}{\fs} pulses directly from a picosecond seed \cite{Sidorenko2019}. It employs a dynamically-tailored gain profile to guide the pulse toward a nonlinear attractor state, thereby minimizing its sensitivity to seed conditions. While the attractor state provides robust performance, it inherently limits the pulse energy to specific fiber parameters, effectively confining the system to a regime that is challenging to scale beyond \SIadj{2}{\micro\joule} energy, even with large-core rod-type fiber \cite{Zhang2021b}. As will be discussed below, pre-chirped amplification that generates sub-\SIadj{100}{\fs} pulses in a \ce{Yb}-doped fiber can be understood as incomplete gain-managed nonlinear amplification with a chirped seed pulse.

This paper serves three purposes: (1) It provides a comprehensive numerical investigation of nonlinear amplification techniques. We find that an analysis that ignores the spectral dependence of amplifier gain provides the main features of nonlinear amplification in an insightful way. The conceptual understanding is still applicable when realistic gain parameters \cite{Lindberg2016,Chen2023} are subsequently included, and this approach helps highlight the important role of the gain spectrum in nonlinear amplification. Moreover, we find that the ``optimal'' nonlinear amplification approach for each seed duration is independent of the chirp of the seed pulse. This allows different regimes of nonlinear amplification to be viewed in a unified manner, which helps delineate and understand them. (2) It describes the discovery and initial experimental demonstration of a new regime based on nonlinear parabolic amplification, but distinct from the self-similar and parabolic pre-shaping approaches. Partial parabolic amplification (PPA) can produce pulses with higher peak power than currently-known nonlinear amplification approaches, while maintaining pulse duration as short as \SI{45}{\fs}. With a photonic crystal fiber (PCF) rod amplifier, scaling up to \SIadj{200}{\MW} peak power is predicted. This would approach the peak power of now-common fiber CPA systems that supply \SIadj{300}{\fs} pulses, and would be an increase of more than \num{6} times over what has been achieved in \SIadj{50}{\fs} fiber amplifiers \cite{Zaouter2008,Zhao2014,Liu2016b,Song2017,Zhang2020a,Zhang2021b} (see Supplementary Sec.~8 \cite{Wang2026}). (3) The role of gain management in rare-earth-doped fiber amplifiers is analyzed in depth, and it is found to play an major role in enabling peak-power scaling in nonlinear amplifiers.

\section{Nonlinear amplification with spectrally-flat gain}
\begin{figure*}[!ht]
\centering
\includegraphics[width=\linewidth]{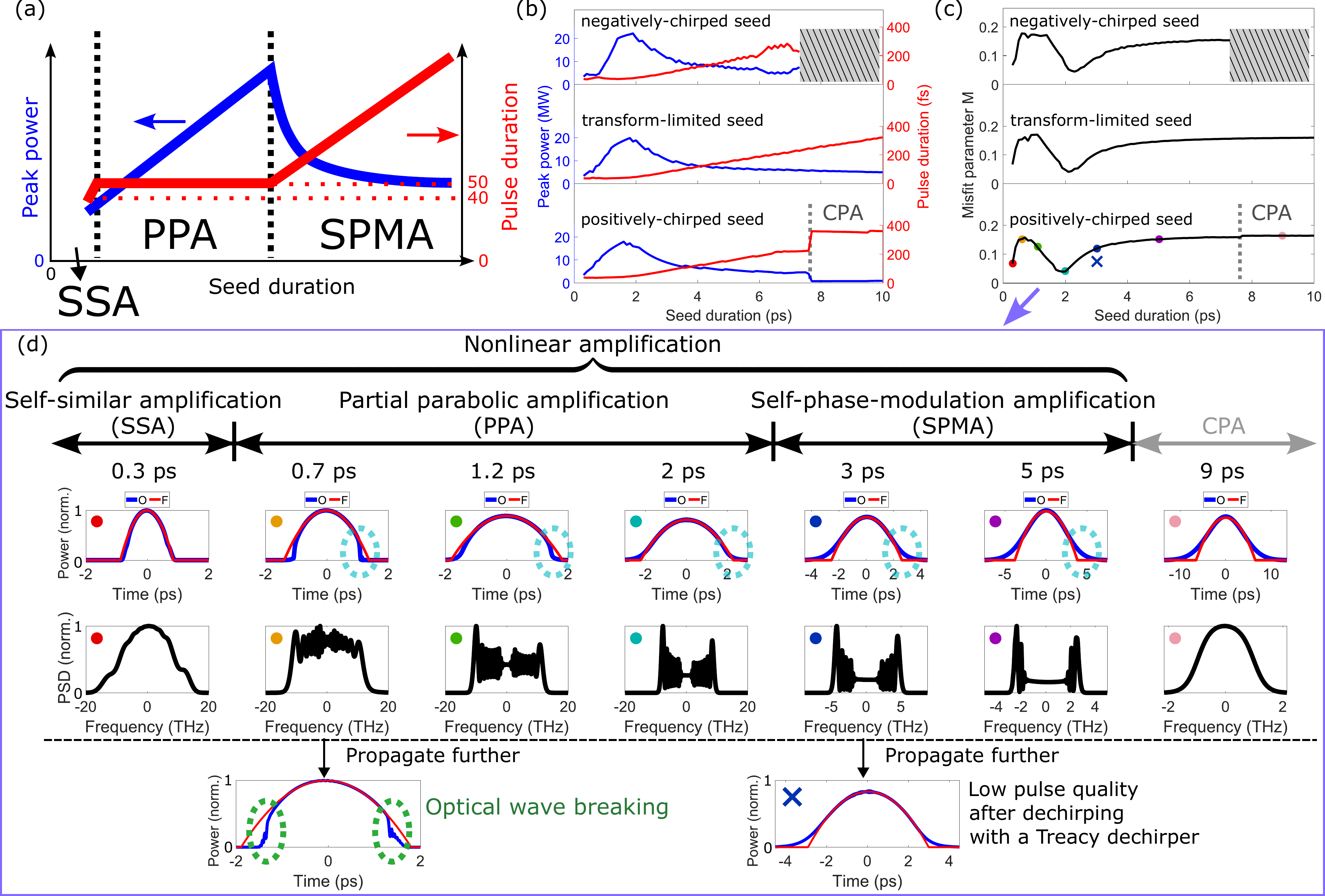}
\caption{Optimal nonlinear amplification regimes at different seed durations, assuming frequency-independent gain. (a) Peak power and duration of the dechirped amplified pulse at different seed durations, with optimal nonlinear amplification regimes indicated. (b) Optimization results for negatively-chirped, transform-limited, and positively-chirped seeds. (c) Misfit parameter $M$ corresponding to results in (b). (d) Temporal and spectral profiles of the amplified pulses for the indicated durations of a positively-chirped seed. Spectra are plotted with respect to frequency relative to the center frequency. O: output temporal profile; F: parabola fitted to the output profile.}
\label{fig:regime_Gaussian}
\end{figure*}

The parameters of the seed pulse -- particularly, the duration -- play a major role in determining the regime in which a nonlinear amplifier operates. For instance, parabolic pre-shaping has been demonstrated using long pulses, but this yields longer compressed durations and consequently lower peak powers \cite{Pierrot2013,Fu2017}. In contrast, self-similar and gain-managed nonlinear regimes achieve sub-\SIadj{100}{\fs} durations, but they require sub-picosecond seed pulses to undergo the desired nonlinear evolution to their respective pulsed states. The short sub-picosecond evolutions restrict the amount of energy that can be stored. On the other hand, a longer seed pulse undergoes an evolution dominated by SPM, which prevents access to nonlinear regimes based on balance of dispersion, gain, and nonlinearity \cite{Finot2006}. Whether there is a nonlinear amplification regime at longer seed durations that is also capable of achieving sub-\SIadj{100}{\fs} compressed durations remains an open question.

To simplify the discussion and establish a foundation for illustrating the role of the gain spectrum, we first assume that the gain is frequency-independent. Analysis of different amplification regimes under this assumption allows isolation of the effects that determine the limiting performance. In the next section, the results of inclusion of an accurate gain model are presented.

\subsection{Optimal nonlinear amplification regimes}
Here, we conduct a numerical investigation of nonlinear amplification in a fiber amplifier, with an exhaustive optimization process over a three-dimensional parameter space (pump power, energy of a Gaussian seed pulse, and incident angle on a standard Treacy grating dechirper \cite{Treacy1969}) at each seed duration [Figs.~\ref{fig:regime_Gaussian}(a,b)]. The optimization procedure is employed to identify the configuration that yields the maximum peak power with a Strehl ratio (defined as the peak power of the dechirped pulse divided by the peak power of the transform-limited pulse) above \num{0.7} and minimal temporal structures, to avoid interference from stimulated Raman scattering (SRS). Numerical simulations employ the unidirectional pulse propagation equation for modeling broadband fields \cite{Chen2023} with refractive index of silica derived from the Sellmeier equation \cite{Malitson1965}. The gain is assumed to be spectrally-flat and saturating, $g=\left(\frac{g_0}{2}\right)\big/\left(1+\frac{E}{E_{\text{sat}}}\right)$, where $E$ is the pulse energy and $E_{\text{sat}}=\frac{h\nu_0A_{\text{eff}}t_{\text{rep}}}{\sigma\tau}$ is the gain saturation energy ($\nu_0$ is the pulse center frequency, $A_{\text{eff}}$ is effective mode area, $\sigma$ is the transition cross section, $\tau$ is the upper-state lifetime of the doped ions, and $t_{\text{rep}}$ is the pulse repetition period). $g_0$ is varied to simulate the variation of pump power, while $E_{\text{sat}}=\SI{164}{\nano\joule}$ is calculated and fixed, assuming a \SIadj{1}{\MHz} pulse train at \SI{1060}{\nm} in a \ce{Yb}-doped fiber with \SIadj{25}{\micro\m} core diameter. The angle of incidence on a \SI{1000}{\text{line}/\mm} Treacy grating dechirper is also varied to accommodate different ratios of cubic-to-quadratic spectral phase. The pulse duration is varied either by imposing negative or positive chirp on a \SIadj{0.3}{\ps} pulse, or by launching a transform-limited pulse with different durations. A double-grating \"{O}ffner configuration is assumed for positively-chirped temporal stretching to ensure perfect compression with a Treacy dechirper \cite{Liu2020}. The gain fiber length is fixed at \SI{1}{\m}. Results of the optimization are shown in Fig.~\ref{fig:regime_Gaussian}, which we will discuss in detail in the following paragraphs.

\subsubsection{Self-similar amplification}
When the seed pulse is short (\SI{0.3}{\ps}), it can rapidly evolve to the self-similar regime \cite{Finot2006,Wabnitz2007}, where it has minimal spectral fringes due to the parabolic temporal profile \cite{Finot2018}. Although the assumed flat gain eliminates the gain-narrowing effect, the resulting compressed duration remains limited to \SI{37}{\fs}. Due to the parabolic temporal profile characteristic of similaritons, the accumulated nonlinear phase is purely quadratic. Consequently, the overall phase cannot be fully compensated by a grating pair, which inherently introduces higher-order dispersion, dominated by TOD. The incompatibility of cubic-to-quadratic phase ratio between the pulse and the dechirper becomes increasingly pronounced as the similariton propagates over extended distances and generates broad bandwidth. As a result, the maximum achievable peak power in self-similar evolution is constrained by TOD.

As the seed pulse duration increases (\SI{\leq0.5}{\ps}), the influence of dispersion during nonlinear evolution diminishes, which results in parabolic shaping confined primarily to the high-power central portion of the pulse [Fig.~\ref{fig:regime_Gaussian}(d)]. There, pronounced spectral broadening occurs, which re-introduces dispersion effects to the overall pulse shaping. To quantify the deviation from an ideal parabolic profile, a misfit parameter $M$ is introduced \cite{Finot2006}:
\begin{equation}
M=\sqrt{\frac{\displaystyle\int\left[\abs{A(t)}^2-\abs{A_{\text{fitted\_parabola}}(t)}^2\right]^2\diff t}{\displaystyle\int\abs{A(t)}^4\diff t}},
\end{equation}
where $A(t)$ and $A_{\text{fitted\_parabola}}(t)$ are temporal profiles (with units of \si{\sqrt{\W}}) of the output pulse and its least-squares fitted parabola [Fig.~\ref{fig:regime_Gaussian}(c)]. Despite the nonlinear shaping of the self-similar regime, the misfit parameter increases, indicating reduced parabolic fidelity of the overall pulse. Nevertheless, the pulse can be compressed well because the deviation from a parabola is limited to a small portion of the temporal edges, and its energy can be further scaled. Limited by the TOD of the dechirper, the spectral bandwidth remains constrained, yielding the same minimum compressed duration of \SI{37}{\fs}. Because a longer pulse must accumulate a larger nonlinear phase to spectrally broaden to the same bandwidth, pulse energy increases with seed duration, which directly enhances the achievable peak power after dechirping.

\subsubsection{Partial parabolic amplification}
Separately from the dynamical shaping through self-similar evolution, deviations from a parabola can also be ``passively'' mitigated by employing a longer seed pulse. Previously-mismatched temporal edges no longer require strong nonlinear shaping, as they are automatically contained by the parabola defined by the central region of the pulse. In cyan dash lines of Fig.~\ref{fig:regime_Gaussian}(d), the parabolically-shaped pulse exhibits temporal edges that deviate from the fitted parabola, extending from below it to above it with increased seed duration. Hence, the misfit parameter drops with increasing duration, reaching a minimum at \SI{2}{\ps} that is determined by the achievable bandwidth in the central asymptotic parabolic evolution [Fig.~\ref{fig:regime_Gaussian}(c)]; the pulse with a larger bandwidth requires a correspondingly-longer duration to counteract the increased dispersive stretching. Further pulse propagation in this regime does not mitigate the deviation, as dispersion is strong enough to induce optical wave breaking [Fig.~\ref{fig:regime_Gaussian}(d)] \cite{Tomlinson1985,Anderson1992,Finot2008}. Eventually, as in the self-similar regime, the performance is limited by TOD, to a compressed duration of \SI{45}{\fs}. This leads to a linear increase of peak power and pulse energy with seed duration. Among regimes of nonlinear amplification, this one ultimately generates pulses with the highest peak power. We refer to it as ``partial parabolic amplification,'' as it generates a partially-parabolic profile during the amplification while maintaining compressibility. It can be distinguished from the self-similar regime, with its dynamical pulse-shaping, and from the parabolic pre-shaping technique, which has no nonlinear shaping during amplification. While parabolic pre-shaping is completely decoupled from the self-similar parameters, partial parabolic amplification preserves the effect of nonlinear parabolic shaping but only around the pulse's peak. This partial decoupling allows PPA to retain the substantial pulse compression of self-similar amplification while extending the parabolic pulse formation to longer duration. The resulting increase in energy extraction capacity yields peak powers that surpass those attainable within the strictly self-similar regime.

Because the propagation does not involve the gain spectral profile, a pulse can evolve to the partial parabolic profile in passive fibers. An interesting application is to use this shaping for pulse compression at the microjoule energy level, beyond what has been previously demonstrated with self-similar \cite{Suedmeyer2003} and SPM \cite{Nikolaus1983} evolutions in solid-core fibers. An attempt was made to exploit a similar evolution in a PCF rod with gain, and this produced \SIadj{5.2}{\micro\joule} and \SIadj{98}{\fs} pulses, limited by the gain spectrum \cite{Saraceno2011}. Supplementary Sec.~2 \cite{Wang2026} describes experimental compression of \SIadj{300}{\fs} pulses to \SI{70}{\fs} with energy of \SI{6}{\micro\joule}, by stretching the input pulses to \SI{2}{\ps} before they traverse a passive photonic crystal fiber and a grating pair. The \SIadj{60}{\MW} peak power of the compressed pulses is the highest produced by a compressor based on a single-mode solid-core fiber, to our knowledge. Moreover, the use of passive fiber enables pulse compression for seeds of arbitrary wavelengths.

\subsubsection{Self-phase-modulation amplification}
If the seed duration exceeds the threshold of partial parabolic shaping, dispersion effect becomes too weak to interact with the nonlinear effect for any parabolic shaping within a practical propagation distance. Under such conditions, the pulse largely preserves its initial temporal shape throughout amplification, with $M$ approaching \num{0.1647} for an ideal Gaussian pulse as the injected seed here [Fig.~\ref{fig:regime_Gaussian}(c)]. Extended propagation may improve the parabolicity of the pulse, but excess accumulated nonlinear phase significantly distorts the dechirped pulse [blue cross in Figs.~\ref{fig:regime_Gaussian}(c,d)]. This regime is thus dominated by the SPM and can be interpreted as a hybrid process of SPM-based nonlinear pulse compression \cite{Tomlinson1984} and amplification. Although SPM generally indicates nonlinear self-phase modulation that underlies all nonlinear amplification techniques, here we restrict its meaning to evolutions that produce deep spectral fringes when the pulse's temporal profile is not specifically controlled \cite{Finot2018}. In contrast to self-similar and partial parabolic amplifications that are constrained by TOD, this SPM amplification (SPMA) regime is bounded by the onset of SRS, resulting in an upper bound of accumulated nonlinear phase. Pulses of different durations are all amplified to the same peak power during the propagation, with longer seed durations corresponding to higher pulse energies. The limited nonlinear phase in longer pulses results in reduced spectral broadening, yielding longer compressed durations. Overall, this regime produces compressed pulses with a linearly-varying duration and a constant peak power with respect to the seed duration [Fig.~\ref{fig:regime_Gaussian}(a)]. Moreover, the narrowing spectrum in this regime reduces the impact of TOD, which contributes to the pulse evolution being constrained by SRS rather than TOD.

It is worth noting that, as evidenced by numerous successful demonstrations of nonlinear post-compression \cite{Nisoli1997,Giguere2009,Koettig2020}, pulses with sufficiently-smooth temporal profiles (\eg a Gaussian pulse) to generate a dominant quadratic nonlinear phase can naturally evolve into the SPM amplification regime. Accordingly, parabolic pre-shaping can be identified as SPM amplification; the constant pulse compression ratio of \num{\sim32} in the transform-limited case [Fig.~\ref{fig:regime_Gaussian}(b)] aligns with prior parabolic pre-shaping experiments (\num{33} and \num{35} times for \SIadj{9}{\ps} \cite{Fu2017} and \SIadj{27}{\ps} \cite{Pierrot2013} pulses, respectively). This shows that parabolic pre-shaping offers no advantage for nonlinear amplification in this regime.

\subsubsection{Optimal amplification regimes at long durations}
Further increasing the seed-pulse duration (\SI{>7.4}{\ps}) reveals distinct amplification dynamics that depend sensitively on the imposed chirp [Figs.~\ref{fig:regime_Gaussian}(b,c)]. For a positively-chirped seed, the pulse evolution abruptly departs from the nonlinear amplification regime and transitions into linear CPA behavior in order to limit the accumulated nonlinear phase, whose cubic-to-quadratic phase ratio is generally incompatible with the Treacy dechirper. For a negatively-chirped seed, effective dechirping would require normal dispersion; under the assumed Treacy configuration, no linear-amplification regime exists. Moreover, no nonlinear amplification regime exists due to an increased uncompensated phase for a longer pulse due to mismatch cubic-to-quadratic phase ratios. In contrast to these two chirping configurations, a transform-limited seed can sustain SPM-dominated amplification even at long pulse durations.

\subsection{Chirp-sign independence of optimal amplification regimes}
Despite differences at long durations, these nonlinear amplification regimes are independent of different chirp signs of the seed at shorter durations [Figs.~\ref{fig:regime_Gaussian}(b,c)]. In the self-similar regime, the pulse evolves toward an asymptotic similariton determined by the pulse energy and fiber parameters \cite{Kruglov2002}, suppressing the influence of minor variations in the initial chirp \cite{Finot2006}. In the partial parabolic and SPM amplification regimes, the substantial nonlinear phase accumulation (\num{10}--\num{40\pi}) dominates the initial chirp. Despite the overall independence of the optimal regimes, it is important to note that the seed parameters to reach these optimal regimes can depend on the initial chirp. As \etal{Finot} pointed out, pulses of different chirp signs lead to distinct types of spectral modulations \cite{Finot2018}. Therefore, the output spectra can be slightly different as well, but the overall bandwidths and peak powers are similar. A negatively-chirped seed pulse can yield slightly-better performance, consistent with studies of pre-chirped amplification \cite{Chen2012,Zhao2014,Liu2016b,Song2017,Luo2018,Chang2019,Zhang2020a,Zhang2021b}. However, prior demonstrations of \ce{Yb}-based sub-\SIadj{100}{\fs} pre-chirped amplification, which keep the input pulses in the \SIadj{<1}{\ps} range, can be understood as imperfect evolution to the GMNA regime, which is an extension of the SPMA regime and is realizable only with a dynamically-varying gain profile; this will be discussed below. Supplementary Secs.~3 and 8 discuss the origin of improvement due to pre-chirping and re-classify some prior works on nonlinear amplification based on the conclusions of this article \cite{Wang2026}.

\subsection{Summary of optimization assuming spectrally-flat gain}
Numerical investigation of amplification with frequency-independent gain reveals three fundamental nonlinear amplification regimes -- self-similar amplification, partial parabolic amplification, and SPM amplification -- each exhibiting distinct temporal and spectral evolution. In the SSA regime, the entire pulse undergoes strong nonlinear evolution and asymptotically approaches the well-known parabolic self-similar state, producing a smooth, fringeless parabolic spectrum. In contrast, SPMA exhibits essentially no temporal parabolic shaping during propagation and develops the characteristic SPM spectrum with deep spectral fringes. PPA serves as an intermediate regime: only the temporal peak experiences strong nonlinear shaping, producing a locally parabolic structure, while the wings remain weakly distorted. Spectrally, PPA inherits features of both SSA and SPMA, exhibiting fringes that are present but shallower than in the SPMA regime.

Despite their differing physical evolution, these regimes share several amplification trends. Both SSA and PPA generate compressed pulses below \SI{50}{\fs} and exhibit peak powers that increase linearly with seed-pulse duration. SPMA, by contrast, produces pulses with nearly constant peak power while the compressed duration and output energy increase linearly with seed duration. The scalability of peak power in SSA and PPA is primarily limited by the TOD of the Treacy dechirper, whereas SPMA is fundamentally constrained by SRS.

Among the three regimes, PPA yields the highest compressed-pulse peak power. This arises from its ability to generate a short \SIadj{45}{\fs} compressed duration while amplifying at a longer seed duration, enabling greater energy extraction. PPA thus combines the broadband, linear-chirp characteristics of SSA with the high-energy storage of SPMA, making it the most favorable regime for maximizing peak power under frequency-independent gain.

It is important to keep in mind that the above conclusions are reached by neglecting the gain spectrum. However, we will see that the physical insight carries over to analysis that includes an accurate treatment of the gain in \ce{Yb}-doped fibers. The gain spectrum modifies some phenomena, and new phenomena can occur. That is the subject of the next section.

\section{Nonlinear amplification in \texorpdfstring{Y\lowercase{b}}{Yb}-doped fiber} 
The results above motivate the experimental demonstration of partial parabolic amplification, which is expected to achieve the maximum peak power along with \SIadj{45}{\fs} pulse duration. However, the gain spectra of rare-earth-doped fiber amplifiers exhibit pronounced frequency dependence, which is further modulated by gain-saturation effects. In this section, we will discuss the effect of frequency-dependent gain and the resulting optimal nonlinear amplifications, which deviate from the flat-gain scenarios. After presenting the optimal amplification regimes in the presence of a frequency-dependent gain, we will discuss its impact on the pulse parameters and describe the results of implementing the PPA regime with gain management.

\subsection{Effect of gain bandwidth}
\begin{figure}[!ht]
\centering
\includegraphics[width=\linewidth]{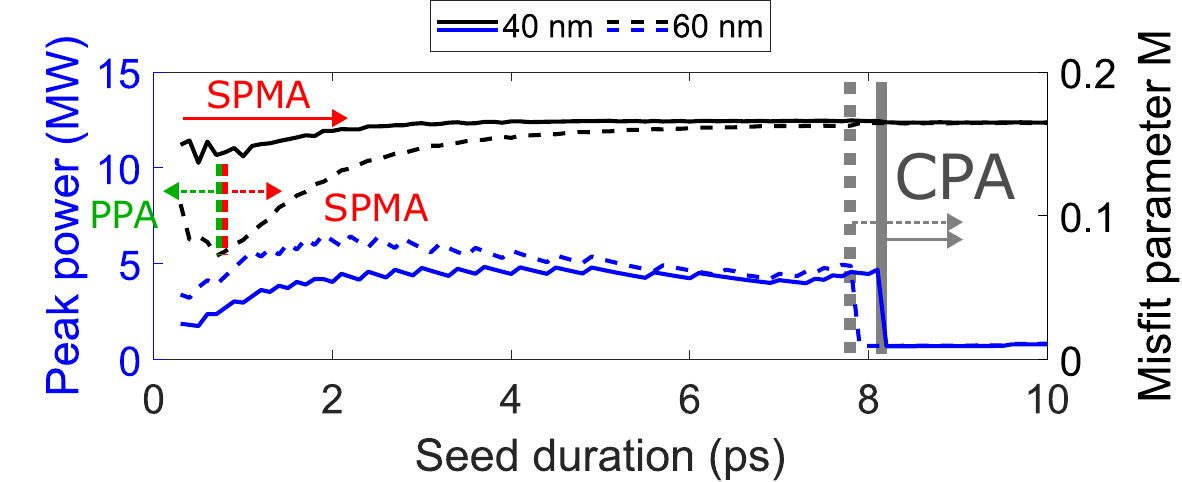}
\caption{Peak power and misfit parameter $M$ of optimal nonlinear amplification regimes with \SIadj{40}{\nm} and \SIadj{60}{\nm} (full-width at half-maximum) Gaussian gain bandwidth.}
\label{fig:regime_Gaussian_narrowband}
\end{figure}

To get an idea of the impact of frequency-dependent gain before performing the analysis with the realistic gain spectrum, we first examine the influence of a Gaussian gain spectrum on the amplification dynamics (Fig.~\ref{fig:regime_Gaussian_narrowband}). Because a (chirped) similariton with narrower bandwidth exhibits a shorter duration \cite{Kruglov2002}, partial parabolic shaping reaches its optimum at a shorter duration under a finite gain bandwidth. With \SIadj{60}{\nm} gain bandwidth, partial parabolic amplification transitions into SPM amplification at \SI{0.7}{\ps}, while there is only SPM amplification regime with \SIadj{40}{\nm} gain bandwidth. Moreover, with these bandwidths, the self-similar evolution does not show up as the optimal regime because the pulse remains too short to store sufficient energy before the spectrum reaches the gain bandwidth.

\subsection{Optimal nonlinear amplification regimes}
Optimization of the peak power of a \ce{Yb}-doped fiber amplifier was performed with an approach similar to that used for the flat-gain case. Numerical simulations assume a \ce{Yb}-doped fiber with \SIadj{25}{\micro\m} core diameter and \SIadj{250}{\micro\m} cladding diameter, with absorption of \SI{2.3}{\dB/\m} at \SI{920}{\nm} (parameters of Thorlabs YB1200-25/250DC-PM). The fiber is co-pumped at \SI{976}{\nm}. The gain model is based on realistic rate equations \cite{Lindberg2016,Chen2023}, and a \SIadj{1}{\MHz} repetition rate is assumed. Since \ce{Yb} has a prominent gain peak at \SI{1030}{\nm} and a smooth gain profile beyond \SI{1050}{\nm} [Fig.~\ref{fig:evolutions}(a)] \cite{Schimpf2010}, we study amplification with seeds at both \SI{1030}{\nm} and \SI{1060}{\nm}. To suppress amplified spontaneous emission (ASE), the gain (ratio of output to input pulse energies) is constrained to be below \num{1000} and \num{300} for \SIadj{1030}{\nm} and \SIadj{1060}{\nm} seeds, respectively. Separate simulations show that ASE amounts to less than a few percent of the output field under these conditions. Because the GMNA regime requires extended fiber lengths to achieve dynamic gain saturation \cite{Sidorenko2019}, the fiber length is allowed to vary, which results in a four-dimensional optimization problem. As the optimization is employed for pulses with longer durations (\SI{\geq1}{\ps}), we find that optimal performance entails shorter fiber lengths, where the SPM or partial parabolic regime is an optimum. Fiber lengths for longer seed durations were fixed to \SI{0.6}{\m} and \SI{0.8}{\m} for \SIadj{1030}{\nm} and \SIadj{1060}{\nm} seeds, respectively. They represent the boundary of the optimal regime with maximum gain. For example, separate examinations with shorter \SI{0.6}{\m} for \SIadj{1060}{\nm} seeds were also conducted and demonstrate higher peak powers for increased seed durations; however, the gain drops to \numrange[range-phrase=--]{20}{250}. If the system reaches the transition to the linear CPA regime, the fiber length is retained from the previous nonlinear amplification regime (\SI{0.6}{\m} and \SI{0.8}{\m} for \SIadj{1030}{\nm} and \SIadj{1060}{\nm} seeds, respectively, as previously mentioned); otherwise, the fiber length continues to decreases to minimize the accumulated nonlinear phase with a sacrifice of gain or pumping efficiency.

\begin{figure*}[!ht]
\centering
\includegraphics[width=\linewidth]{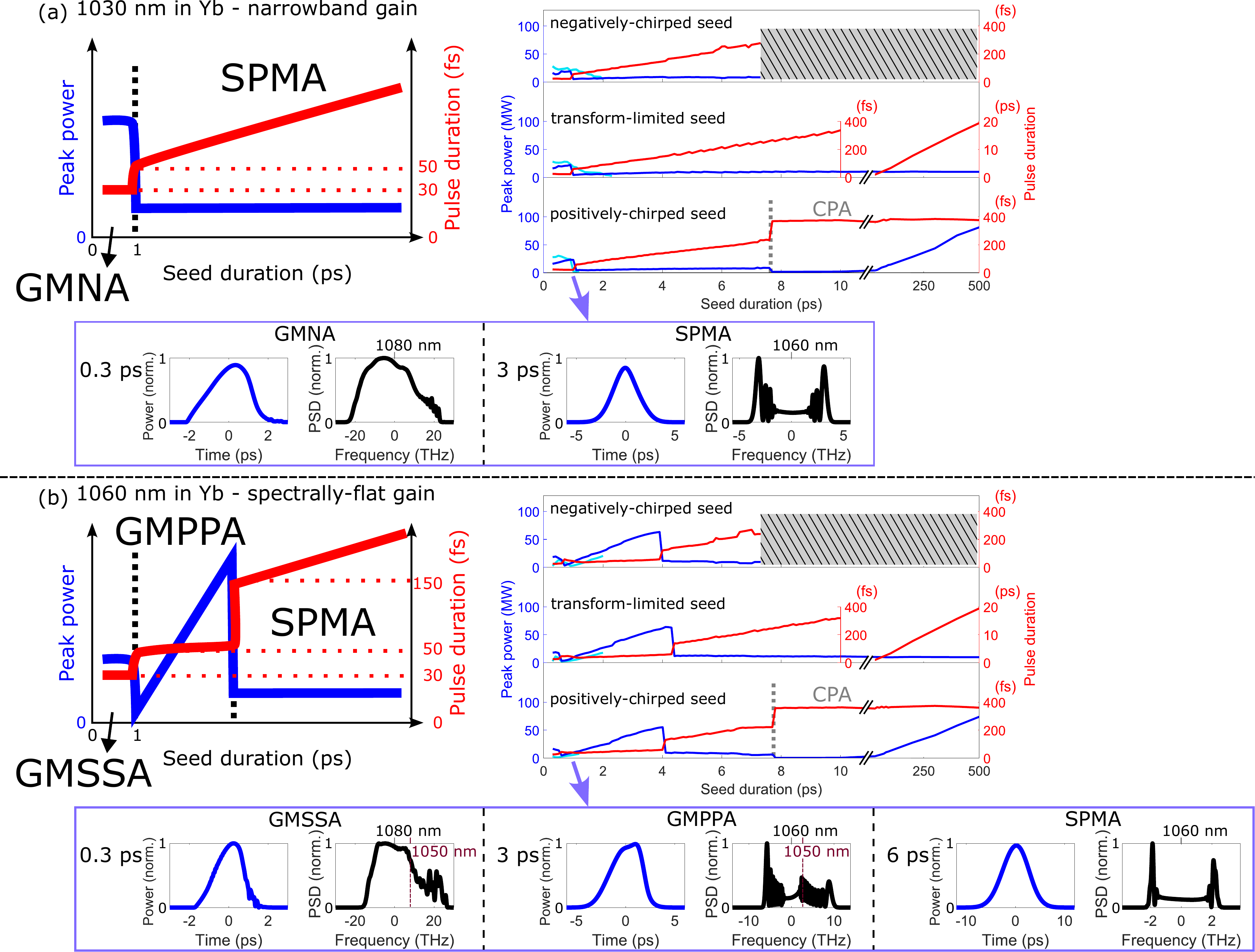}
\caption{Optimal nonlinear amplification regimes at different seed durations and chirp signs in a \SIadj{25}{\micro\m}-core \ce{Yb}-doped fiber for \SIadj{1030}{\nm} and \SIadj{1060}{\nm} seeds. Optimization with half the doping concentration is also conducted at short durations for optimal GMNA; resulting peak powers are shown with light blue lines. The \SIadj{1050}{\nm} long-wavelength edge of the \ce{Yb} high-gain region is labeled in the bottom spectra. In addition, we label \SI{1080}{\nm} for GMSSA because it exhibits significant spectral broadening up to \SI{1150}{\nm}, as in GMNA \cite{Sidorenko2019}.}
\label{fig:regime_Yb}
\end{figure*}

To streamline the discussion of various amplification regimes, below we will use SSA for self-similar amplification, PPA for partial parabolic amplification, and SPMA for SPM amplification, as well as GMNA for gain-managed nonlinear amplification and CPA for chirped-pulse amplification.

Prior to examining the specifics of nonlinear amplification in \ce{Yb}, we first outline the relevant amplification regimes associated with it (Fig.~\ref{fig:regime_Yb}). For a \SIadj{1030}{\nm} seed, only GMNA and SPMA regimes exist [Fig.~\ref{fig:regime_Yb}(a)]. Because \ce{Yb} exhibits a relatively-narrow \SIadj{\sim40}{\nm} gain bandwidth at \SI{1030}{\nm} [Fig.~\ref{fig:evolutions}(a)] \cite{Schimpf2010}, in principle, only the SPMA regime emerges as the optimal nonlinear amplification (Fig.~\ref{fig:regime_Gaussian_narrowband}). However, the dynamically-varying gain spectrum allows the pulse to nonlinearly evolve into the GMNA regime beyond what was discussed in the flat-gain scenario. For a \SIadj{1060}{\nm} seed, the amplification regimes are similar to those found with flat gain (Fig.~\ref{fig:regime_Gaussian}). The major difference lies in the emergence of gain management in both the SSA and PPA regimes. Gain management is found to play an important role in increasing the peak power in gain-managed SSA (GMSSA), as well as extending the PPA regime to longer seed durations and thus higher peak powers, which we refer to as gain-managed PPA (GMPPA). In the next paragraph, we will discuss in detail how gain management in rare-earth-doped fibers enhances nonlinear amplification, even beyond what is achievable with spectrally-flat gain.

\subsection{Gain management}
\begin{figure}[!ht]
\centering
\includegraphics[width=\linewidth]{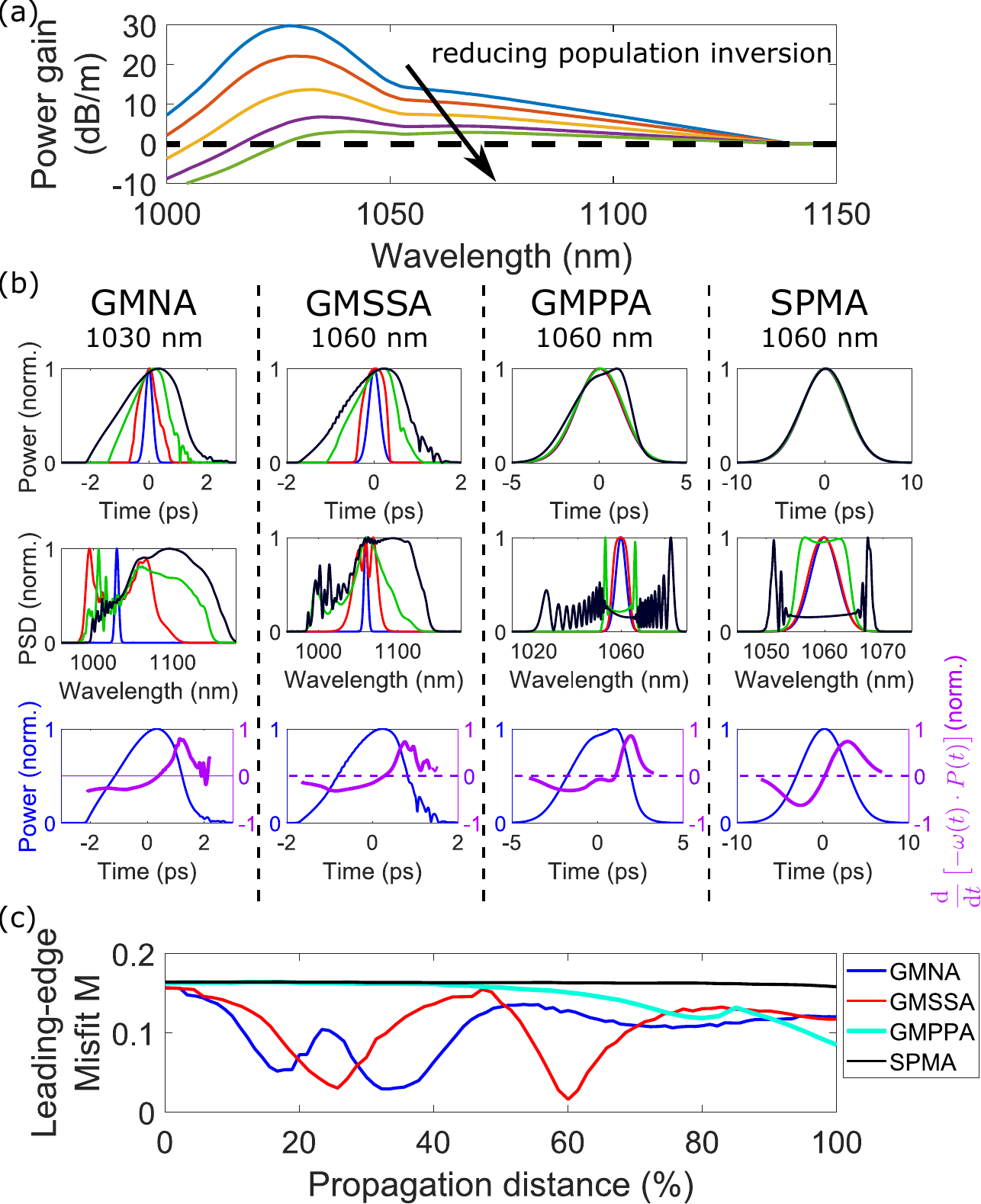}
\caption{Illustration of gain management for each regime in \ce{Yb}-doped fiber. (a) \ce{Yb} gain spectrum with \SIadj{976}{\nm} pumping. (b, top and middle rows) Temporal and spectral evolutions of each nonlinear amplification regimes. Evolutions are shown by fields at \num{0}, \num{33}, \num{67}, and \SI{100}{\percent} propagation distances with blue, red, green, and black lines, respectively. (b, bottom row) Induced frequency shift of the output profiles, from nonlinear chirp generation [Eq.~(\ref{eq:freq_shift})]. The effect of TOD compensation from gain management occurs in GMNA, GMSSA, and GMPPA, while an effective cubic phase is not produced in SPMA. (c) Leading-edge misfit parameter $M$ of different regimes. To exclude the effect of gain-managed distortion at short wavelengths, $M$ is computed based only on the leading edge of the temporal profile, which corresponds to the long-wavelength part of the spectrum that stays in the flat-gain region. Variation of $M$ in GMNA and GMSSA result from the imperfectly-flat long-wavelength gain, which creates an extended leading temporal edge as shown in (b). Only pulses in the SPMA regime do not undergo effective parabolic shaping.}
\label{fig:evolutions}
\end{figure}

Gain management in \ce{Yb} shows several distinct behaviors compared to the idealized case of constant flat gain [Fig.~\ref{fig:evolutions}(a)]. As the inversion decreases, the gain spectrum transitions from net gain to net loss below \SI{1030}{\nm}, where a broad region of spectrally-flat gain emerges, spanning approximately \num{1030} to \SI{1140}{\nm} \cite{Schimpf2010}. It is important to emphasize that the applicability of the flat-gain scenarios discussed earlier (Fig.~\ref{fig:regime_Gaussian}) is governed by the degree of ``spectral flatness'' rather than the absolute bandwidth because the asymptotic self-similar evolution is fundamentally defined under conditions of spectrally-flat gain \cite{Fermann2000,Kruglov2002}. Despite the limited spectral extent in the long-wavelength region (the gain is reasonably high from \SI{1050}{\nm} to \SI{\sim1100}{\nm}), which avoids the pronounced gain narrowing near \SI{1030}{\nm}, the gain remains sufficiently flat to support effective SSA and PPA pulse evolutions.

\begin{figure}[!ht]
\centering
\includegraphics[width=\linewidth]{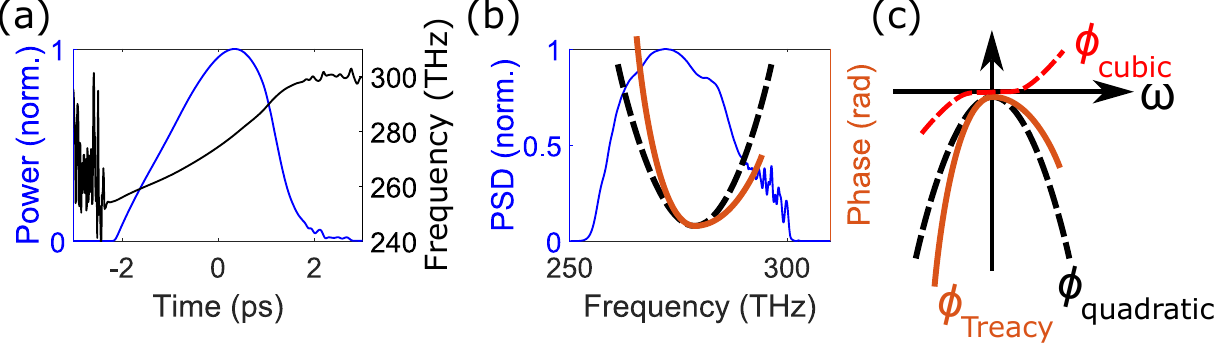}
\caption{Illustration of TOD compensation induced by gain management. (a) Temporal profile (blue) and instantaneous frequency (black), along with (b) spectral profile (blue) and its phase (orange), of a GMN pulse. Phase deviation from a parabola (black dash line) is exaggerated for visualization purposes. (c) Phase of a Treacy dechirper, where $\phi_{\text{Treacy}}=\phi_{\text{quadratic}}+\phi_{\text{cubic}}$. The GMN temporal profile induces a slow leading redshifting and fast trailing blueshifting during pulse propagation, effectively leading to significant temporal stretching of low-frequency components in the leading edge but less stretching for high-frequency parts in the trailing edge. This means that spectral components of lower frequency have larger spectral phases than expected from a parabola, while components of higher frequency have smaller spectral phases, as shown in the orange line and black dash line in (b). Treacy dechirper provides negative group delay dispersion with positive TOD, which cancels the spectral phase of the GMN pulse.}
\label{fig:TOD_phase}
\end{figure}

Gain management enhances nonlinear amplification differently for different regimes (Fig.~\ref{fig:evolutions}). In GMNA seeded at \SI{1030}{\nm}, the pulse initially evolves in the SPM regime (in SPMA). Due to the spectrally-flat gain beyond \SI{1050}{\nm}, newly-generated spectral components in this region undergo effective self-similar evolution. Those around \SI{1030}{\nm} that arise from excessive nonlinear phase from SPM are absorbed by the gain fiber when they encounter strongly-saturated gain. These processes lead to a parabolic long-wavelength leading edge, and rapidly-falling short-wavelength trailing edge on the pulse. Nonlinear chirp generation follows
\begin{align}
\dod{\omega(t)}{z} & =\dod{}{z}\left(-\dod{\phi_{\text{NL}}(t)}{t}\right) \nonumber \\
& =\dod{}{z}\left(-\dod{}{t}\int\gamma(\omega(t))\cdot P(t)\diff z\right) \nonumber \\
& =\dod{}{t}\left[-\gamma(\omega(t))\cdot P(t)\right]\propto\dod{}{t}\left[-\omega(t)\cdot P(t)\right] \nonumber \\
& \hspace{10.7em}\approx-\omega_0\dod{}{t}P(t),
\label{eq:freq_shift}
\end{align}
where $\phi_{\text{NL}}(t)$ is the accumulated nonlinear phase, $\gamma(\omega(t))\propto\omega(t)$ is the nonlinear coefficient, $\omega(t)$ is the instantaneous frequency and is almost constant due to the smooth and relatively-small extent of spectral broadening compared to the pulse's center frequency $\omega_0$, and $P(t)$ is the instantaneous power. Hence, the temporal profile with slowly-rising leading edge and rapidly-falling trailing edge leads to slow nonlinear redshifting at the leading edge but fast blueshifting at the trailing edge [purple lines in the bottom row of Fig.~\ref{fig:evolutions}(b)]. This effectively induces a negative cubic spectral phase that can be compensated by a Treacy dechirper (Fig.~\ref{fig:TOD_phase}). In this way, GMNA alleviates the TOD constraint found in the flat-gain scenario, which enables operation up to the higher Raman threshold. Pulses amplified in this regime can be dechirped to near the transform limit with a grating pair. Pulses as short as \SI{25}{\fs} have been produced \cite{Buttolph2022}, with peak powers up to \SI{28}{\MW}. Since this process relies on strong gain saturation, it is best realized with lower doping concentrations, to allow the pulse to propagate over extended fiber lengths where dispersion can sufficiently stretch the pulse and enhance energy extraction [light blue lines in Fig.~\ref{fig:regime_Yb}(a)]. Gain-managed amplification in the GMSSA and GMPPA regimes follows a distinct evolution. Initially, pulses propagate within their respective regimes due to their confinement in the \SIrangeadj{1050}{1140}{\nm} flat-gain region. As the pulse spectrally broadens and enters the high-gain spectral region below \SI{1050}{\nm}, short-wavelength components undergo enhanced amplification, which yields a pronounced short-wavelength trailing edge. Meanwhile, the long-wavelength components -- having evolved under SSA or PPA dynamics -- retain a parabolic temporal profile, and form a smooth leading edge. This asymmetric shaping induces a TOD-compensating effect analogous to that observed in GMNA, thereby extending the operational limit toward the Raman threshold. The sharp trailing edge induces significant spectral broadening toward short wavelengths, which generates the broad bandwidth that supports \SIadj{25}{\fs} and \SIadj{45}{\fs} durations for the GMSSA and GMPPA regimes, respectively. Although the compressed duration in the GMPPA regime is the same as that in the flat-gain scenario, this mechanism delays the transition from the high-peak-power (GM)PPA regime to the low-peak-power SPMA regime, from \SI{2}{\ps} in the flat-gain scenario (Fig.~\ref{fig:regime_Gaussian}), to \SI{4}{\ps}. Misfit parameter $M$ is computed based on the leading edge of the temporal profile and confirms that GMNA, GMSSA, and GMPPA experience self-similar parabolic shaping in their long-wavelength parts [Fig.~\ref{fig:evolutions}(c)].

Self-similar evolution occurs in both GMNA and GMSSA (as evidenced by spectra without deep fringes \cite{Finot2018}) but the evolutions differ overall. In GMSSA, the self-similar evolution is an early stage. For GMNA, it emerges later in the process, where it arises from the redshifting gain spectrum. The \SIadj{1030}{\nm} gain region in \ce{Yb} asymmetrically shapes the pulse through absorption in GMNA or through amplification in GMSSA and GMPPA, which lead to the same TOD compensation effect.

Recently, \etal{Turitsyn} presented an interesting simplified model of nonlinear pulse propagation in a medium with spectrally-asymmetric gain (a simple linear dependence on frequency) \cite{Turitsyn2023}. Solutions include temporally-asymmetric pulses that can accumulate large nonlinear phases without wave-breaking, so the model captures key features of GMNA. However, self-similar evolution after the broadened spectrum enters the flat-gain region at long wavelengths is naturally not addressed by this minimal model, nor are the development of a largely-linear chirp and compensation of TOD through gain management. It is also not surprising that the simple linear gain model is not adequate for the GMSSA and GMPPA regimes.

The results summarized in Fig.~\ref{fig:regime_Yb} allow comparison of the performance in different amplification regimes. Among the nonlinear regimes, GMPPA clearly achieves the highest peak power. The highest achievable peak power with GMPPA is comparable to CPA with stretching to \SI{500}{\ps}, while its compressed duration is \num{6} times shorter.

\subsection{Experimental demonstration of GMPPA}
\begin{figure}[!ht]
\centering
\includegraphics[width=\linewidth]{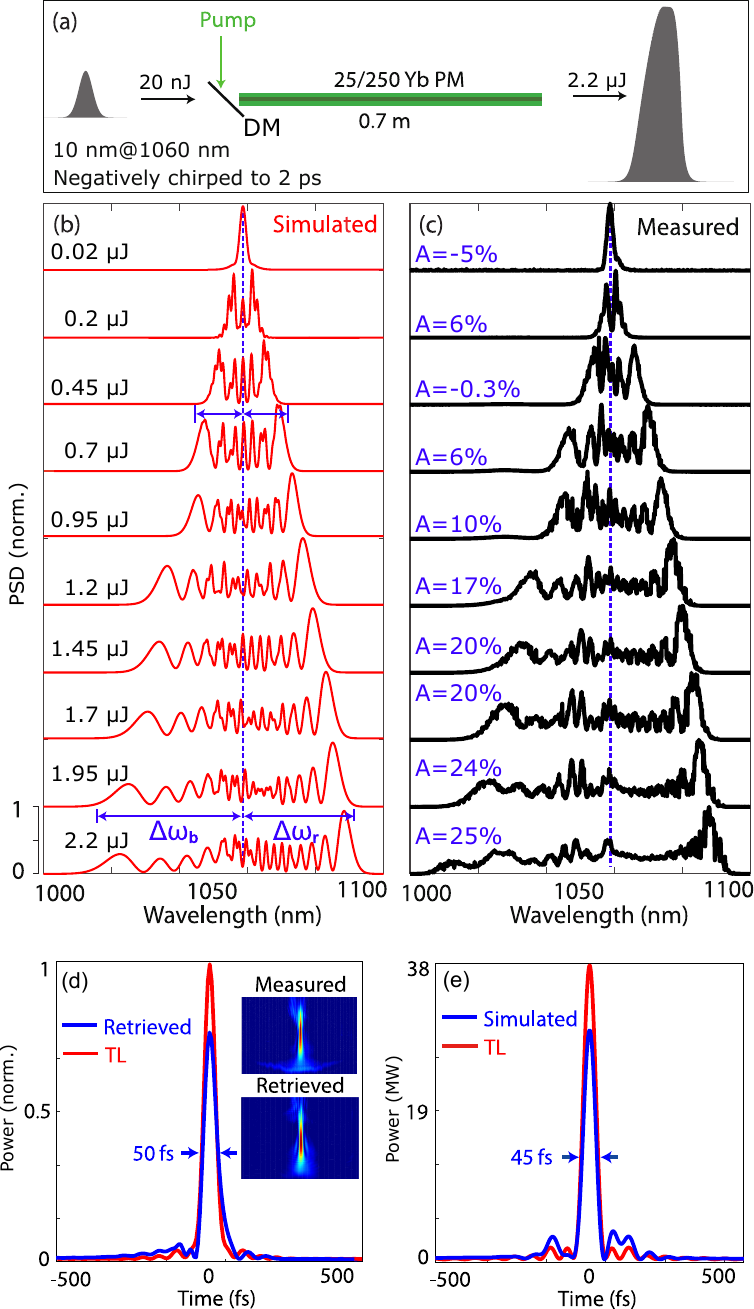}
\caption{Experimental validation of the GMPPA regime. (a) Schematic of the experimental setup. DM: dichroic mirror. (b) Simulated and (c) measured output spectra for pulse energies indicated at left. (d) FROG measurement and (e) simulated dechirped pulse (blue), compared with their transform-limited (TL) pulses (red).}
\label{fig:Measurements}
\end{figure}

Guided by the results above, we investigated GMPPA experimentally [Fig.~\ref{fig:Measurements}(a)]. Since nonlinear amplification is insensitive to the initial chirp, for convenience we applied a negative chirp to our seed pulse with a grating pair. A \SIadj{20}{\nano\joule} seed pulse, negatively chirped to \SI{2}{\pico\second}, was amplified in \SI{0.7}{\meter} of \ce{Yb}-doped double-cladding polarization-maintaining fiber with \SIadj{25}{\micro\m} core diameter and \SIadj{250}{\micro\m} cladding diameter, and absorption of \SI{2.3}{\dB/\m} at \SI{920}{\nm} (Thorlabs YB1200-25/250DC-PM). The fiber was coiled with a diameter of \SI{7.5}{\cm} to operate in the fundamental mode and pumped with up to \SI{30}{\W} at \SI{976}{\nm}. A practical challenge is the provision of sufficiently-energetic seed pulses at \SI{1060}{\nm}, given that the gain is constrained to around \SI{20}{\decibel} to avoid excessive ASE. We produced the seed by spectrally filtering a \SIadj{10}{\nm} band around \SI{1060}{\nm} from a home-built \SIadj{1}{\MHz} and \SIadj{0.5}{\micro\joule} GMNA source, which yielded \SI{60}{\nano\joule}. After losses from an isolator, grating pair for pre-chirping, fiber coupling, and fiber coiling, we recorded \SI{20}{\nano\joule} at the amplifier output with the pump turned off. The amplified pulses were dechirped using a Treacy dechirper and characterized with second-harmonic frequency-resolved optical gating (FROG).

Simulations and experimental results are presented in Figs.~\ref{fig:Measurements}(b) and (c). The simulations were done with the experimental parameters, across a range from \num{0.02} to \SI{2.2}{\micro\joule}, limited by the onset of strong ASE with the available seed-pulse energy. The simulated and measured spectra agree very well at all energies. The output spectra display strong spectral fringes, consistent with results above (Figs.~\ref{fig:regime_Gaussian} and \ref{fig:regime_Yb}). As the energy increases, the spectrum initially broadens symmetrically through PPA. In this regime, PPA resembles SPMA, and can only be identified conclusively by measuring the partial parabolic shaping effect in its temporal profile. When the spectrum reaches the high-gain region below \SI{1050}{\nm}, it broadens asymmetrically, with a rapid extension toward short wavelengths, and the asymmetry distinguishes GMPPA from SPMA. The rapid spectral broadening, which results from the fast-falling trailing edge, counters gain narrowing from the high-gain region, which is consistent with the observations. The \SIadj{2.2}{\micro\joule} pulses have \SIadj{50}{\fs} duration and Strehl ratio of \num{0.77} [Fig.~\ref{fig:Measurements}(d)], in good agreement with simulations [Fig.~\ref{fig:Measurements}(e)]. The peak power reaches \SI{30}{\MW}. Very similar results are obtained with a slightly-different gain fiber, and direct measurements of the peak power confirm the high pulse quality (see Supplementary Sec.~5 \cite{Wang2026}). 

To quantify the effect of gain management, we introduce an asymmetry parameter $A=\left(\triangle\omega_{\text{blue}}-\triangle\omega_{\text{red}}\right)/\left(\triangle\omega_{\text{blue}}+\triangle\omega_{\text{red}}\right)$ [Figs.~\ref{fig:Measurements}(b) and (c)], where $\triangle\omega_{\text{blue}}$ and $\triangle\omega_{\text{red}}$ represent the spectral widths from the initial center frequency to the blue-shifted and red-shifted edges, respectively [Fig.~\ref{fig:Measurements}(b)]. In the initial stage of GMPPA evolution (\ie energies lower than \SI{0.7}{\micro\joule}), the asymmetry parameter $A\approx0$. As the pulse energy increases, the spectrum reaches the high-gain region and experiences asymmetric broadening, so $A>0$. If the seed wavelength lies below \SI{1030}{\nm}, the spectrum approaches the high-gain region from the short-wavelength side, leading to enhanced long-wavelength components of the pulse and thus $A<0$. Unlike the cases discussed previously, this evolution generates a sharply-rising leading edge, and the resulting nonlinear phase accumulation introduces a positive cubic spectral phase that undesirably reinforces the Treacy dechirper’s inherent TOD. Therefore, the sign of the gain-managed TOD is dictated by the direction of spectral asymmetry, which in turn depends on the seed wavelength. A preferential extension toward shorter wavelengths is crucial for gain-managed TOD compensation. Please see Supplementary Sec.~7 for more details \cite{Wang2026}.

\section{Discussion}
GMPPA yields a highly-modulated spectral profile that closely resembles that observed in the SPMA regime. Without the initial study of amplification with flat gain, such pulses may be mischaracterized as merely undergoing SPM. The investigation with flat gain exposes the partial parabolic shaping, which is not clearly visible in a \ce{Yb}-doped fiber amplifier due to later gain-induced distortion. It also lays the groundwork for recognizing the critical role of gain management in shaping pulse dynamics through nonlinear compensation of TOD. The overall analysis suggests that all regimes of nonlinear amplification are now thoroughly characterized, with partial parabolic amplification emerging as the most-promising for the generation of \SIadj{50}{\fs} pulses.

Scaling of the pulse energy and peak power in the GMPPA regime is ultimately limited by the seed-pulse energy. The need to avoid ASE restricts the gain at \SI{1060}{\nm} to about \SI{20}{\decibel}. Reaching the \SIadj{10}{\micro\joule} level will require a \SIadj{100}{\nano\joule} seed pulse. Filtering of a GMN amplifier with the highest energy achievable (\SI{1}{\micro\joule} \cite{Sidorenko2020}) is simple but provides only \SI{60}{\nano\joule} in a \SIadj{10}{\nm} spectral band. A modest CPA system could be used to generate \SI{100}{\nano\joule} at \SI{1060}{\nm} \cite{Verhoef2014,Chang2019,Zhao2022}, for example. Double-pass amplification \cite{Zaouter2011,Zhang2021b,Zhao2022} or fiber-based regenerative amplification \cite{Haig2023a} may be alternative configurations to enable the use of a weak seed, but their evolutions change the gain dynamics discussed here and thus require further investigation. We conclude that for sub-\SIadj{100}{\fs} pulses, GMNA remains the simplest approach below \SIadj{1}{\micro\joule} energy, while GMPPA becomes the solution for higher energies. Recent demonstration of CPA in the broad flat-gain region of \ce{Yb} also shows promise in high-peak-power generation by use of GMNA as a front end \cite{Zhang2025}. At energies beyond \SI{15}{\micro\joule}, GMPPA will be limited by self-focusing even with a large-mode-area fiber \cite{Zhihua2012}, as the peak power in the fiber will be near the critical power. After dechirping, the peak power should reach \SI{200}{\MW}, which would be an order-of-magnitude increase for \SIadj{50}{\fs} fiber amplifiers. For pulse energies above \SI{15}{\micro\joule} or peak powers above \SI{200}{\MW}, a high-energy CPA combined with a pulse compressor will be more suitable. To put these values in context, peak power up to \SI{4}{\GW} has been obtained with a fiber CPA system, but the pulse duration is \SI{120}{\fs} and the system is quite complex \cite{Stark2021}.

GMPPA and GMNA are alone in being able to generate sub-\SIadj{50}{\fs} amplified pulses. In addition to the possibility of scaling to higher performance, GMPPA offers some practical advantages. It can work with a short gain fiber, such as a PCF rod amplifier, whereas GMNA requires fibers longer than existing rod amplifiers to achieve effective dynamical gain saturation (\eg \SI{5}{\m} in \cite{Sidorenko2019} and \SI{2.5}{\m} in \cite{Sidorenko2020}). If desired, gain can be sacrificed to reach the shortest pulse duration, which allows GMPPA to function as a high-energy post-compression stage that delivers \SIadj{50}{\fs} pulses, not only without loss but with gain. Pulses produced by GMPPA accumulate half as much nonlinear phase as pulses produced by GMNA (Supplementary Sec.~1 \cite{Wang2026}), so GMPPA may resist the nonlinear polarization coupling that limits GMNA in non-PM fibers \cite{Chen2023}. Conventional techniques to improve CPA, such as counter-pumping and injection of a circularly-polarized seed, may be explored with GMPPA, but are not compatible with GMNA. With all these combined, GMPPA exhibits a practical limit of self-focusing-limited \SI{15}{\micro\joule}, an order of magnitude higher than the \SIadj{\sim1}{\micro\joule} limit of GMNA imposed by the largest mode size (\SIadj{\sim30}{\micro\m} diameter) achievable in coilable fibers.

This article focuses on identifying the amplification regimes that maximize peak power under the practical constraint of employing a Treacy dechirper. It is possible to operate in sub-optimal conditions. For example, a long picosecond pulse propagating through an extended fiber, longer than required by a short seed pulse, may eventually reach the self-similar regime \cite{Finot2006,Hammani2012}; however, the accumulation of excess quadratic phase will lead to low-quality dechirped pulses due to incompatible cubic-to-quadratic phase ratio with the dechirper. Similarly, an excessively-energetic \SIadj{2}{\ps} seed may undergo significant initial spectral broadening, which suppresses partial parabolic shaping and instead leads to amplification governed by the SPMA regime. A short femtosecond pulse can enter the self‑similar regime under comparatively weak amplification, enabled by a lower doping concentration and compensated by a longer propagation distance; however, the extended interaction exacerbates Raman-induced temporal distortion \cite{Soh2006}. These operations only generate pulses with limited peak powers. Among the nonlinear amplification regimes known to date, GMPPA achieves the maximum peak power.

The Supplement contains discussions of the effects of chirp sign (Supplementary Sec.~3), seed duration (Supplementary Sec.~6), and seed wavelength (Supplementary Sec.~7) on the optimal nonlinear amplification regimes \cite{Wang2026}.

\section{Conclusion and Perspective}
In summary, systematic optimization of the peak-power performance of nonlinear fiber amplifiers allows useful and insightful identification of distinct regimes based on the pulse propagation. Among these, the combination of gain management and nonlinear evolution to a partially-parabolic time profile is found to offer the best performance and its characteristic features are illustrated. Initial experiments with limited seed-pulse energy clearly demonstrate the new regime and its ability to generate \SIadj{50}{\fs} pulses with higher energy than achieved by existing techniques. The crucial role of gain management in the generation of \SIadj{50}{\fs} pulses from amplifiers is elucidated, and the design of such amplifiers is illustrated. With some reasonably-straightforward development, it should be possible to scale GMPPA to the \SIadj{200}{\MW} peak power range in the future, which would help fill a gap in fiber amplifiers between existing nonlinear amplifiers and CPA systems. It will also be interesting to explore gain-managed techniques (GMNA, GMSSA, and GMPPA) with \ce{Er}, \ce{Nd}, and \ce{Tm} fiber gain media, as well as on different amplification platforms, including fiber regenerative amplifiers \cite{Haig2023a} to mitigate the ASE constraint.

\section{Backmatter}
\begin{backmatter}
\bmsection{Funding} National Institutes of Health (R01EB033179, U01NS128660).

\bmsection{Acknowledgments} W.W. acknowledges a Postdoctoral Fellowship from Shanghai Jiao Tong University for partial support of his studies at Cornell University. Y.-H.C. was partially supported by a Mong Fellowship from Cornell Neurotech. The authors express gratitude to Almantas Galvanauskas, Lauren Cooper, and Liang Dong for valuable discussions.

\bmsection{Disclosures} The authors declare no conflicts of interest.

\bmsection{Data availability} The code used in this work has been made publicly available at \url{https://github.com/AaHaHaa/MMTools}. 


\end{backmatter}
\bibliography{reference.bib}

\bibliographyfullrefs{reference.bib}

\end{document}


\maketitle

\newpage 
\section{Optimization procedure}\label{sec:Optimization_procedure}
The optimization relies on a simple local search over discrete variables. The searching direction moves toward the highest peak power. This process is performed for each seed duration with the result from the previous duration as the initial condition for the current duration. If the spacing between durations is small enough, such initial condition should approach the optimum very well. In addition, rather than scanning over a huge parameter space for each duration, we perform many scans, with each scan applying over a small neighboring parameter space. If it finds an optimum at the boundary, this optimum acts as a new center for the new small parameter space in the next scan, until the optimum finally stays at the center of small parameter space. This optimization scheme ensures only local optimum, not global optimum. Therefore, we have also tried to search with different initial conditions to ensure that the optimum we found is sufficiently good. A more-advanced optimization scheme could be the subject of future work.

Below we show optimization results summarized in Figs.~1 and 3 in the article. The angle of incidence on the \SIadj{1000}{\text{line}/\mm} Treacy dechirper can vary between \num{10} and \num{80} degrees to accommodate a wide range of cubic-to-quadratic phase ratios achievable in gratings of different linewidths.

\clearpage
\subsection{Spectrally-flat gain}
\begin{figure}[!ht]
\centering
\includegraphics[width=\linewidth]{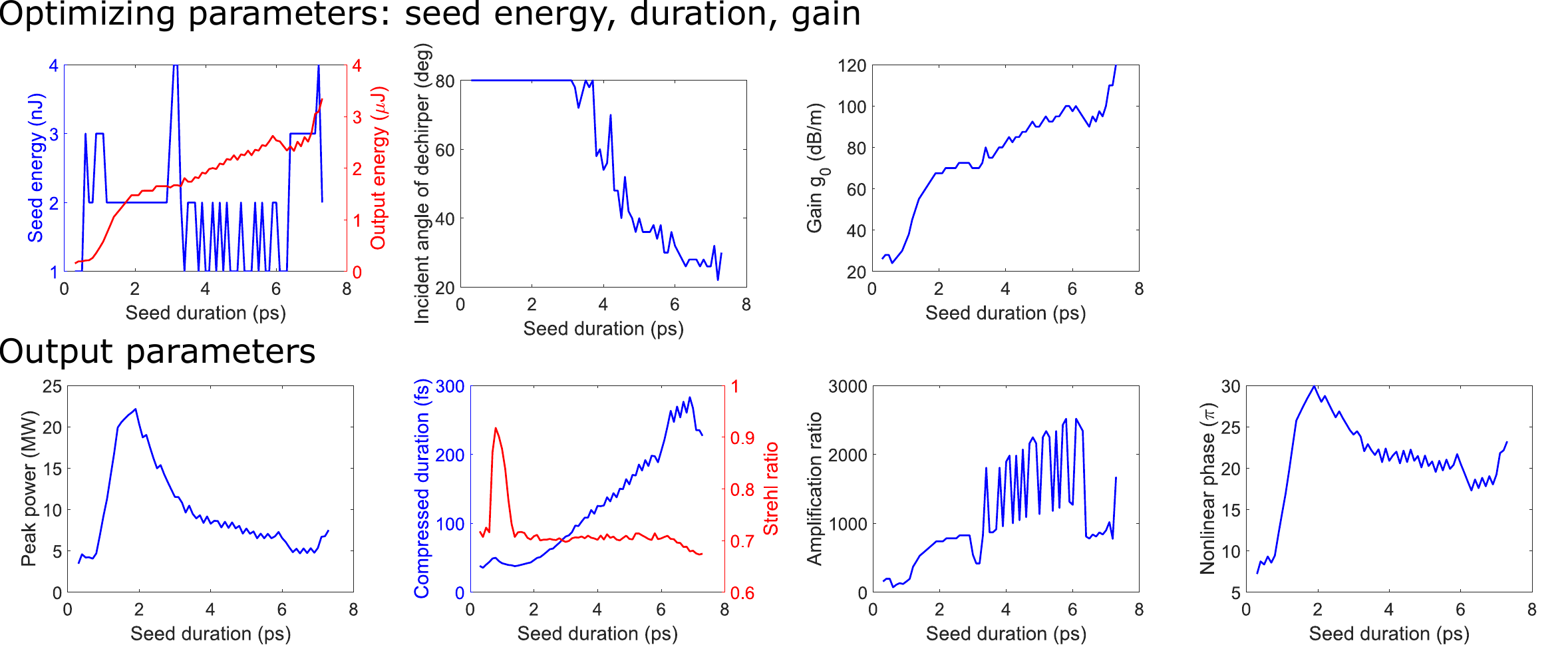}
\caption{Optimization results for \textbf{negatively-chirped} seeds with spectrally-flat gain. The top row shows how the optimal parameters vary with seed-pulse duration, obtained by maximizing the peak power of the dechirped output. The bottom row presents the corresponding parameter values resulting from this optimization.}
\label{fig:flat_negatively_chirped_opt}
\end{figure}

\begin{figure}[!ht]
\centering
\includegraphics[width=\linewidth]{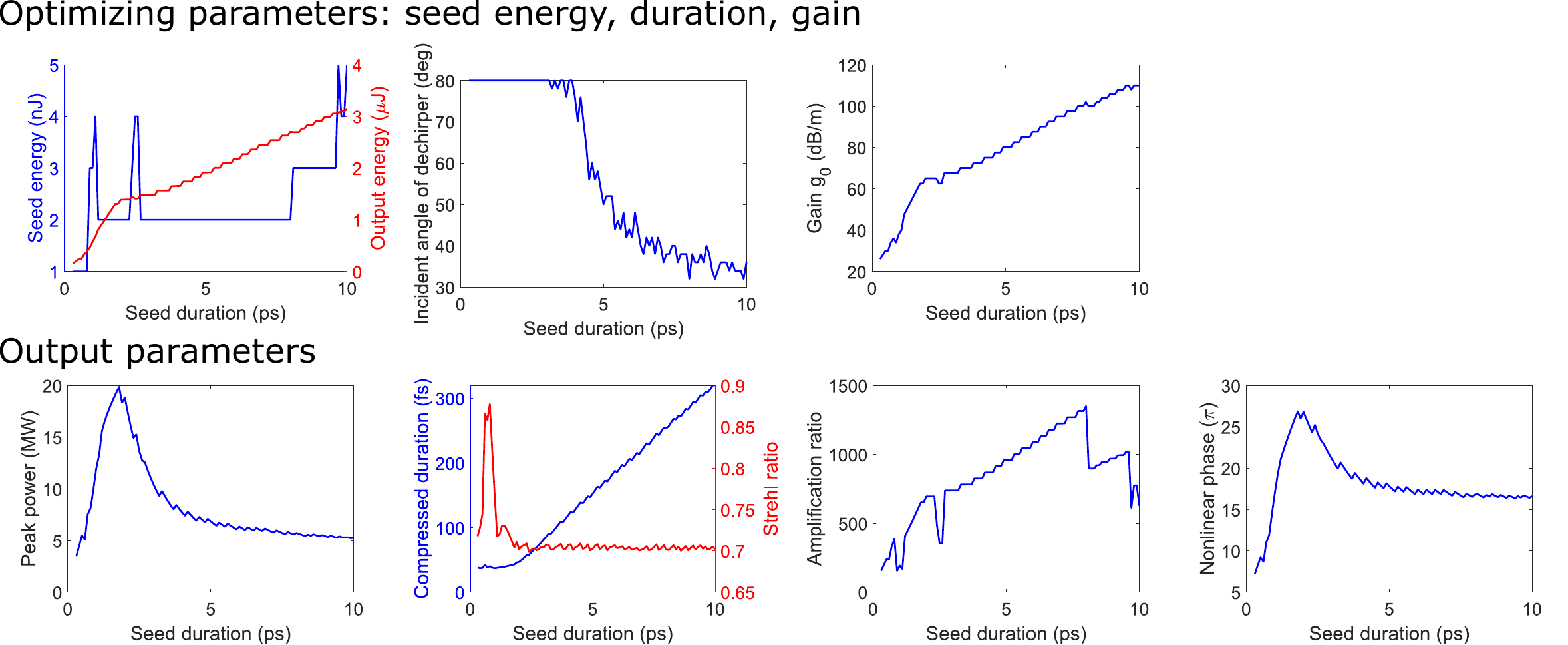}
\caption{Optimization results for \textbf{transform-limited} seeds with spectrally-flat gain. The top row shows how the optimal parameters vary with seed-pulse duration, obtained by maximizing the peak power of the dechirped output. The bottom row presents the corresponding parameter values resulting from this optimization.}
\label{fig:flat_transform_limited_opt}
\end{figure}

\begin{figure}[!ht]
\centering
\includegraphics[width=\linewidth]{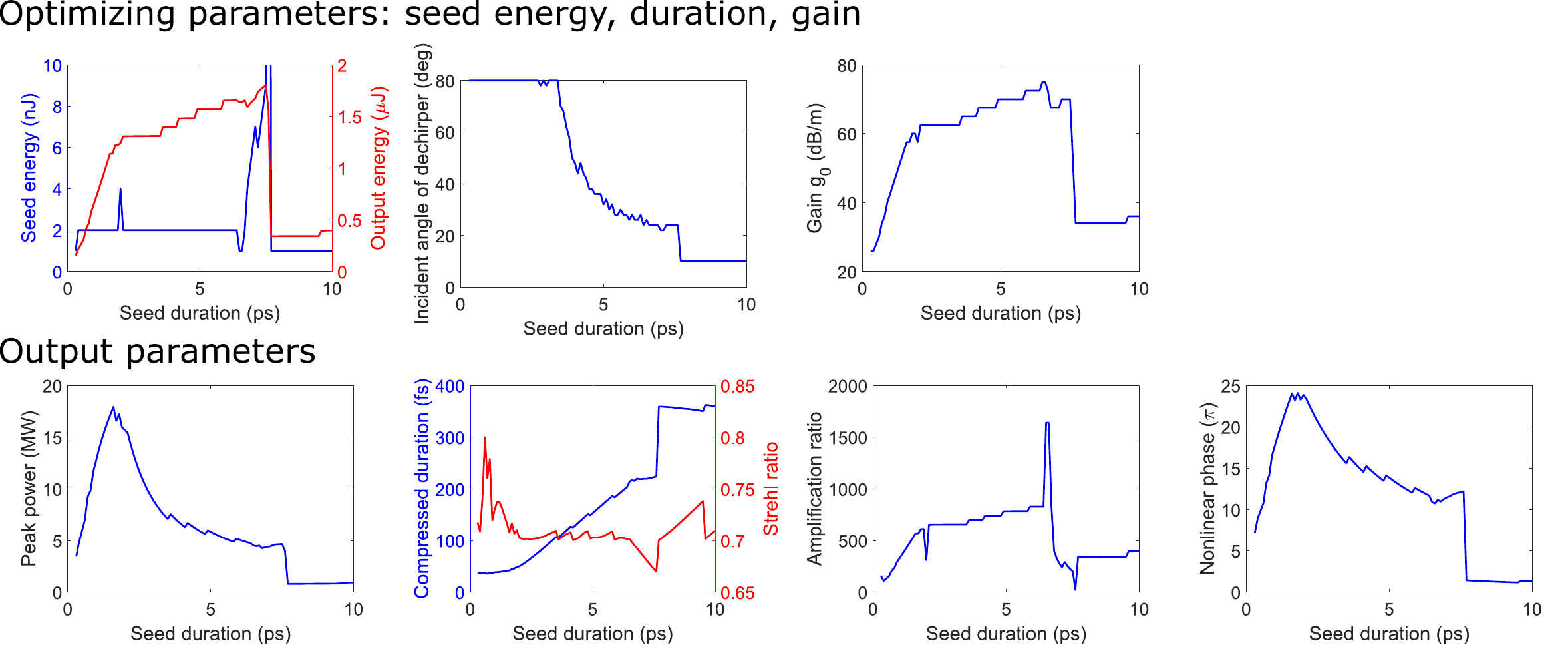}
\caption{Optimization results for \textbf{positively-chirped} seeds with spectrally-flat gain. The top row shows how the optimal parameters vary with seed-pulse duration, obtained by maximizing the peak power of the dechirped output. The bottom row presents the corresponding parameter values resulting from this optimization.}
\label{fig:flat_positively_chirped_opt}
\end{figure}

Beyond the primary conclusions presented in the article, the optimization results (Figs.~\ref{fig:flat_negatively_chirped_opt}--\ref{fig:flat_positively_chirped_opt}) reveal several additional insights.

In the absence of ASE constraint, the optimal seed energy remains approximately invariant with seed duration. The output energy after amplification increases with duration, and the corresponding amplification factor exhibits a similar upward trend. The accumulated nonlinear phase reaches its maximum at the highest-peak-power PPA state near \SI{2}{\ps}, then decreases to an approximately constant value at long durations due to the Raman-imposed limitation characteristic of the SPMA regime.

\begin{figure}[!ht]
\centering
\includegraphics[width=\linewidth]{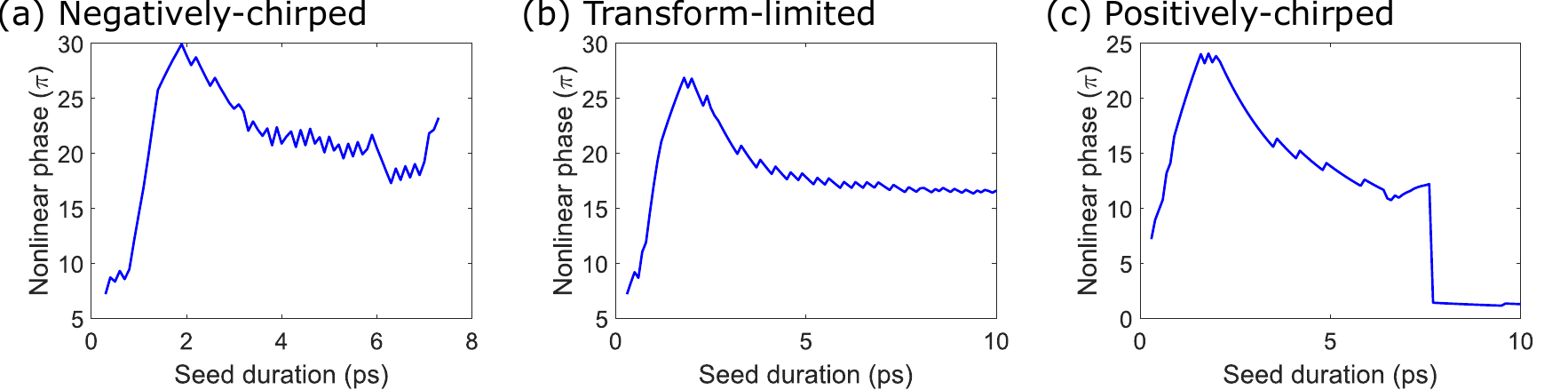}
\caption{Accumulated nonlinear phase for different chirp conditions with spectrally-flat gain.}
\label{fig:Nonlinear_phase_flat}
\end{figure}

\clearpage
\subsection{Yb gain} 
\begin{figure}[!ht]
\centering
\includegraphics[width=\linewidth]{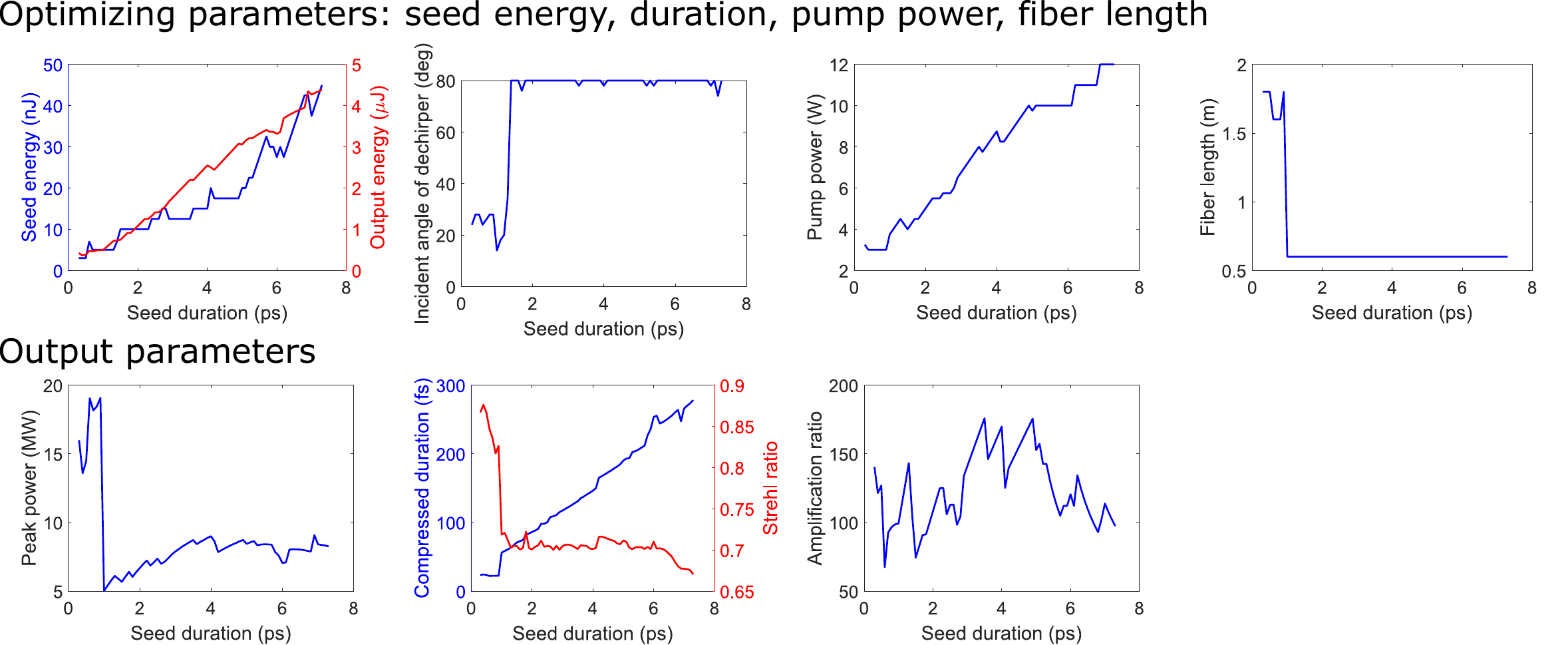}
\caption{Optimization results for \textbf{negatively-chirped} seeds at \textbf{\SI{1030}{\nm}} with \ce{Yb} gain. The top row shows how the optimal parameters vary with seed-pulse duration, obtained by maximizing the peak power of the dechirped output. The bottom row presents the corresponding parameter values resulting from this optimization.}
\label{fig:Yb_negatively_chirped_opt_1030}
\end{figure}

\begin{figure}[!ht]
\centering
\includegraphics[width=\linewidth]{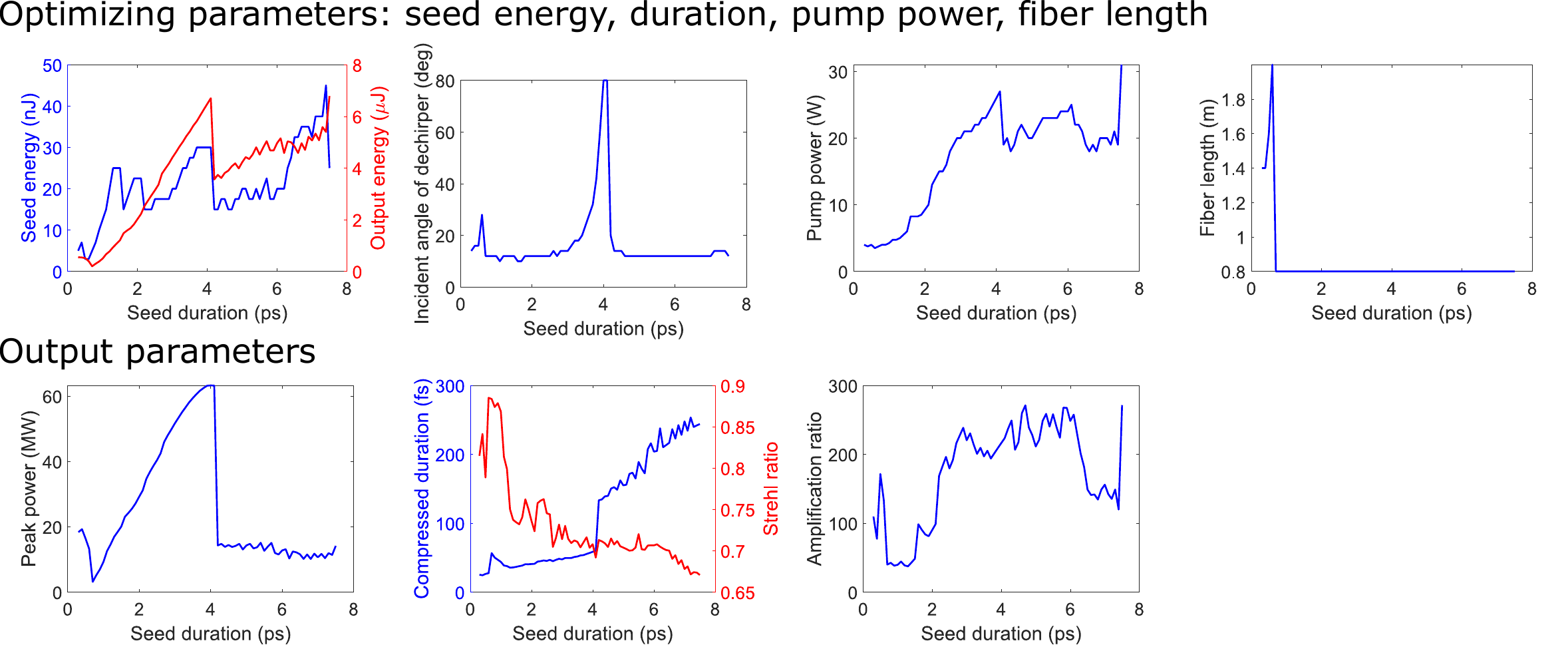}
\caption{Optimization results for \textbf{negatively-chirped} seeds at \textbf{\SI{1060}{\nm}} with \ce{Yb} gain. The top row shows how the optimal parameters vary with seed-pulse duration, obtained by maximizing the peak power of the dechirped output. The bottom row presents the corresponding parameter values resulting from this optimization.}
\label{fig:Yb_negatively_chirped_opt_1060}
\end{figure}

\begin{figure}[!ht]
\centering
\includegraphics[width=\linewidth]{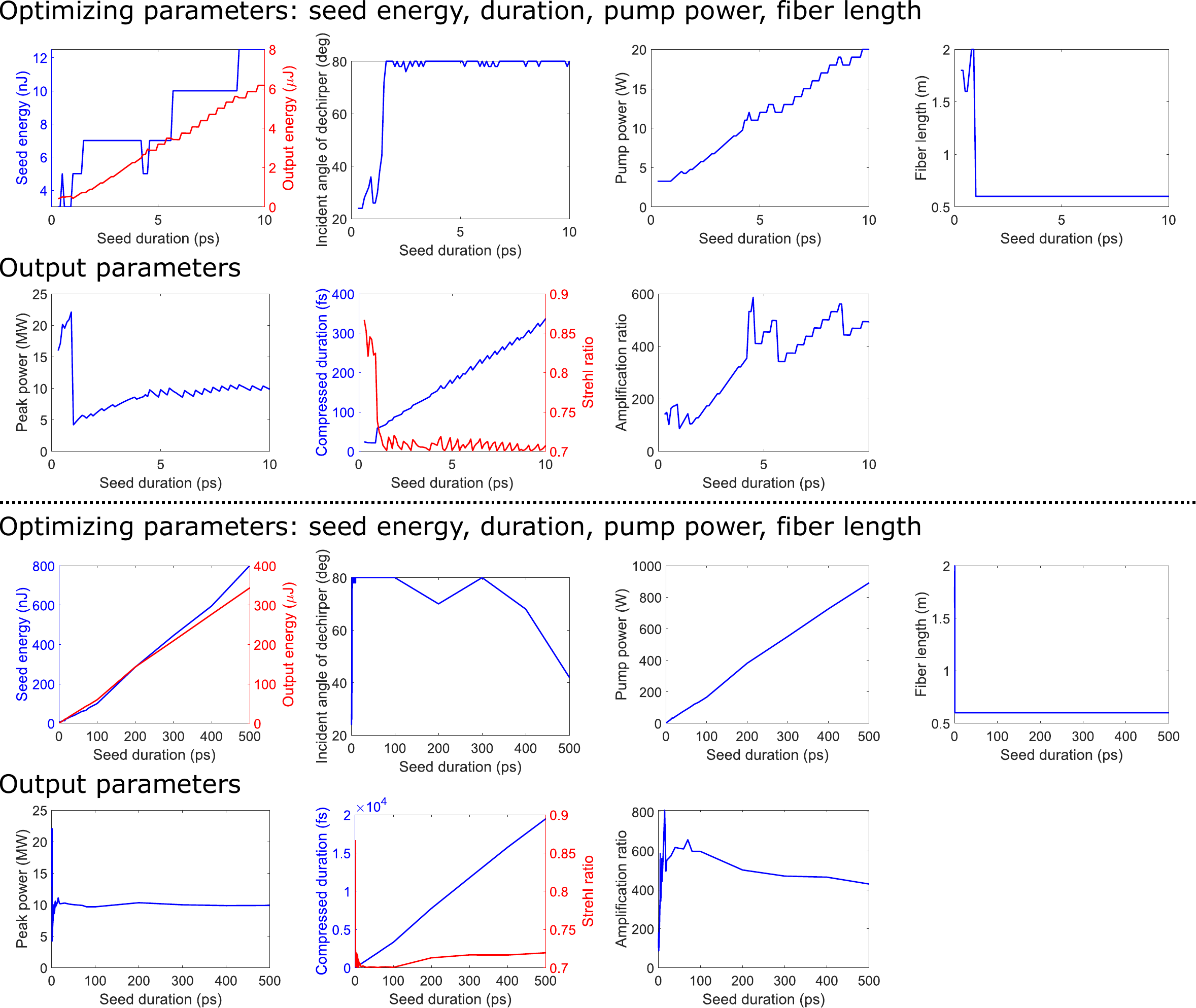}
\caption{Optimization results for \textbf{transform-limited} seeds at \textbf{\SI{1030}{\nm}} with \ce{Yb} gain. There are two panels: the top one covers \SIrangeadj{0.3}{10}{\ps} seed duration; the bottom one covers up to \SI{500}{\ps}. The top row of each panel shows how the optimal parameters vary with seed-pulse duration, obtained by maximizing the peak power of the dechirped output. The bottom row of each panel presents the corresponding parameter values resulting from this optimization.}
\label{fig:Yb_transform_limited_opt_1030}
\end{figure}

\begin{figure}[!ht]
\centering
\includegraphics[width=\linewidth]{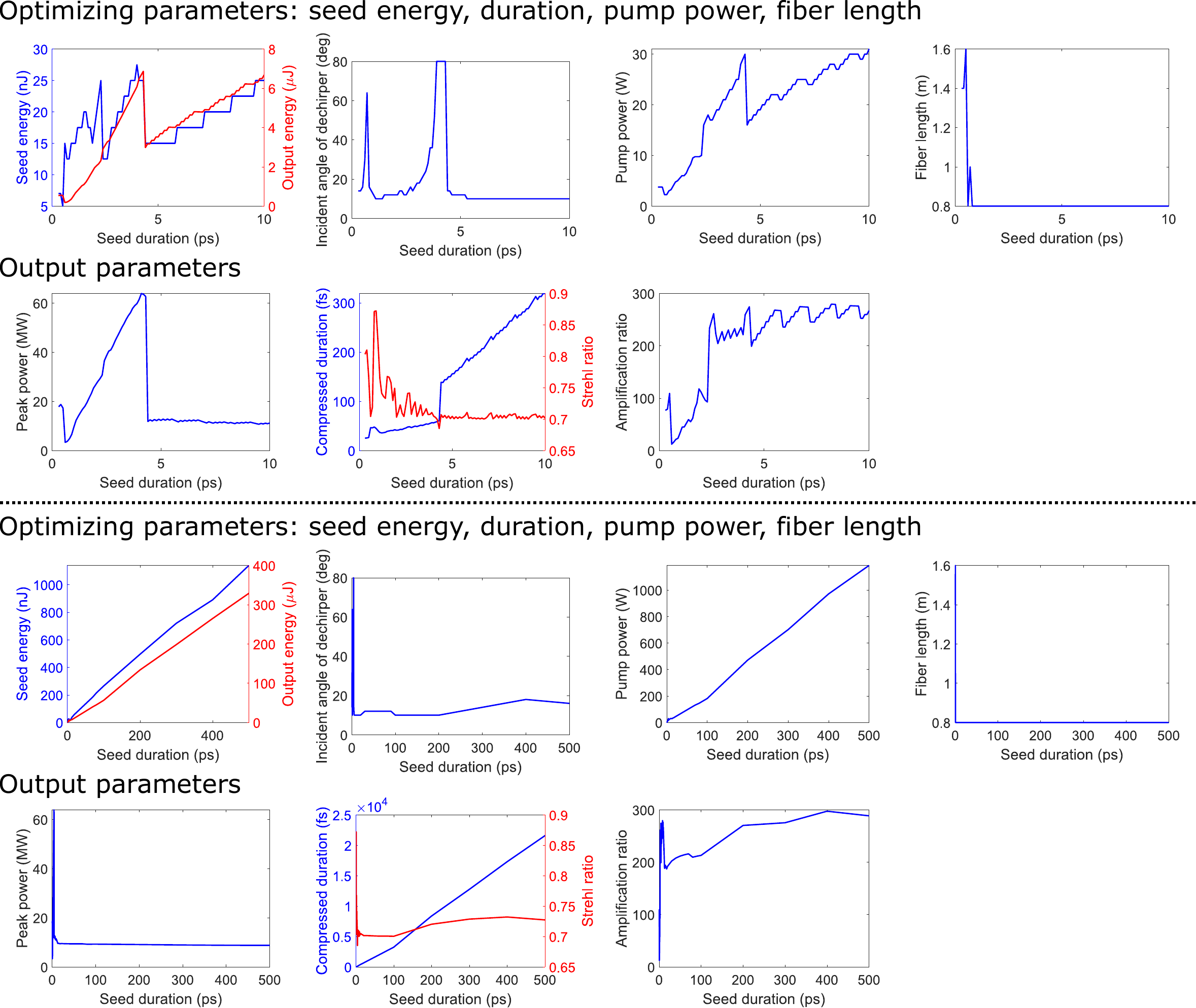}
\caption{Optimization results for \textbf{transform-limited} seeds at \textbf{\SI{1060}{\nm}} with \ce{Yb} gain. There are two panels: the top one covers \SIrangeadj{0.3}{10}{\ps} seed duration; the bottom one covers up to \SI{500}{\ps}. The top row of each panel shows how the optimal parameters vary with seed-pulse duration, obtained by maximizing the peak power of the dechirped output. The bottom row of each panel presents the corresponding parameter values resulting from this optimization.}
\label{fig:Yb_transform_limited_opt_1060}
\end{figure}

\begin{figure}[!ht]
\centering
\includegraphics[width=\linewidth]{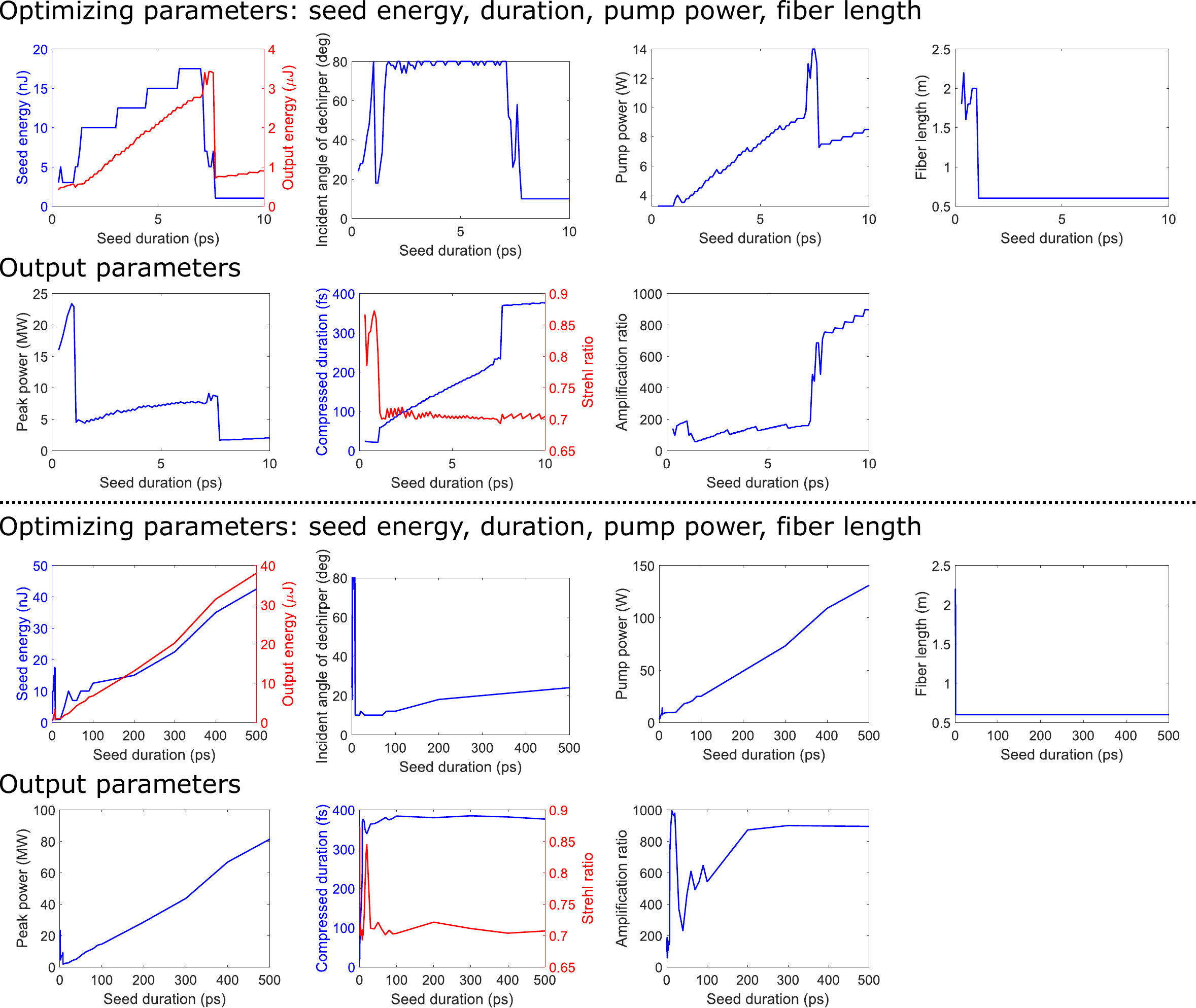}
\caption{Optimization results for \textbf{positively-chirped} seeds at \textbf{\SI{1030}{\nm}} with \ce{Yb} gain. There are two panels: the top one covers \SIrangeadj{0.3}{10}{\ps} seed duration; the bottom one covers up to \SI{500}{\ps}. The top row of each panel shows how the optimal parameters vary with seed-pulse duration, obtained by maximizing the peak power of the dechirped output. The bottom row of each panel presents the corresponding parameter values resulting from this optimization.}
\label{fig:Yb_positively_chirped_opt_1030}
\end{figure}

\begin{figure}[!ht]
\centering
\includegraphics[width=\linewidth]{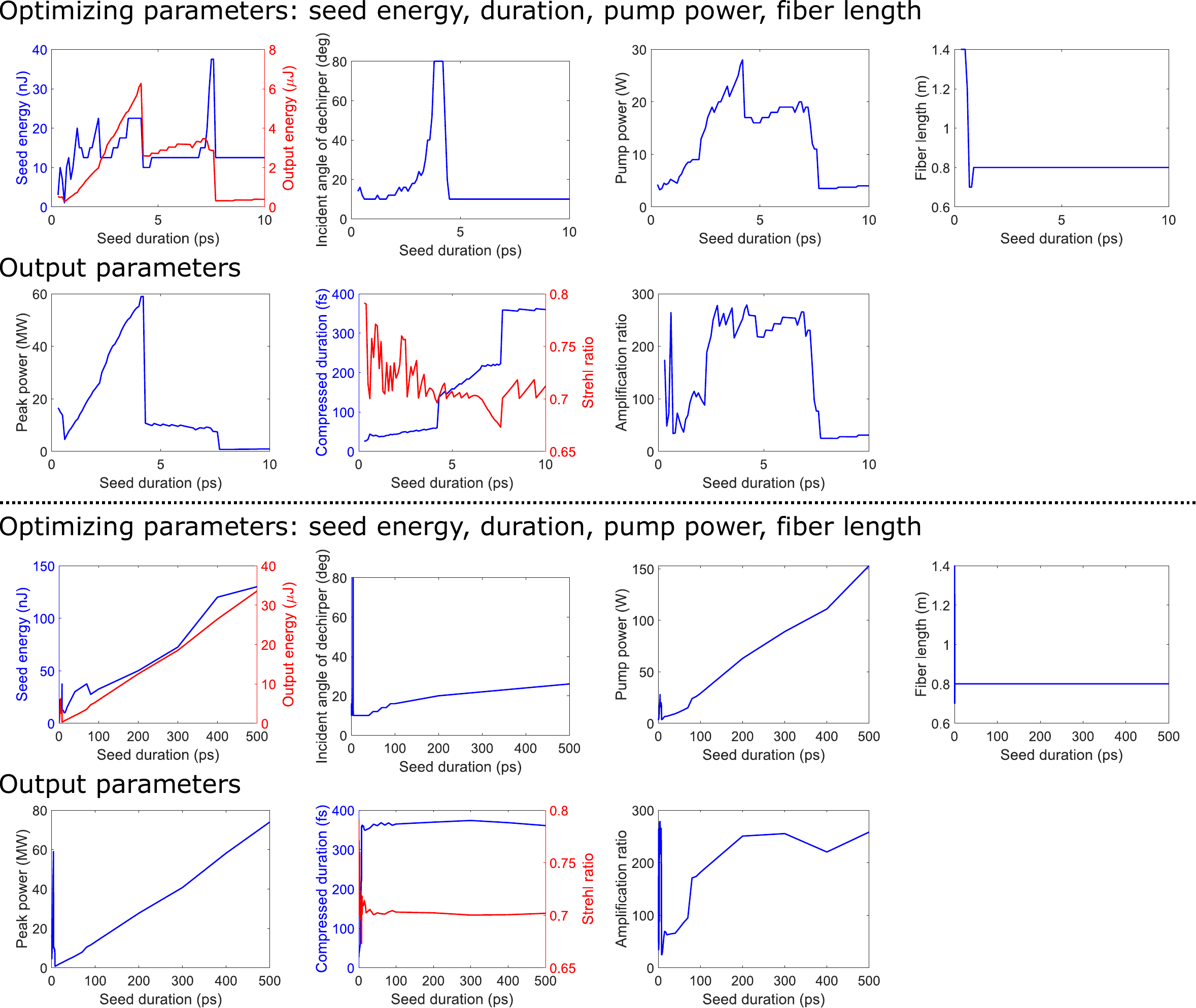}
\caption{Optimization results for \textbf{positively-chirped} seeds at \textbf{\SI{1060}{\nm}} with \ce{Yb} gain. There are two panels: the top one covers \SIrangeadj{0.3}{10}{\ps} seed duration; the bottom one covers up to \SI{500}{\ps}. The top row of each panel shows how the optimal parameters vary with seed-pulse duration, obtained by maximizing the peak power of the dechirped output. The bottom row of each panel presents the corresponding parameter values resulting from this optimization.}
\label{fig:Yb_positively_chirped_opt_1060}
\end{figure}

Beyond the primary conclusions presented in the article, the optimization results (Figs.~\ref{fig:Yb_negatively_chirped_opt_1030}--\ref{fig:Yb_positively_chirped_opt_1060}) reveal several additional insights.

For \SIadj{1030}{\nm} seeds, the optimal seed energy roughly linearly increases with seed duration due to the dominance of the SPMA regime. The output energy after amplification scales linearly too, with a roughly-fixed amplification ratio. Variation trend of pump power follows the output energy.

For \SIadj{1060}{\nm} seeds, the optimal seed energy remains roughly fixed at \SI{20}{\nano\joule} within the GMPPA and SPMA regime with a slight increase with the duration following the output energy, which further roughly fixes the amplification ratio at around \num{200}. The variation of output energy of the GMPPA pulse follows the same trend as the peak power. In the SPMA regime, the peak power is fixed with a linearly-increasing output energy due to a linearly-increasing dechirped duration.


\begin{figure}[!ht]
\centering
\includegraphics[width=\linewidth]{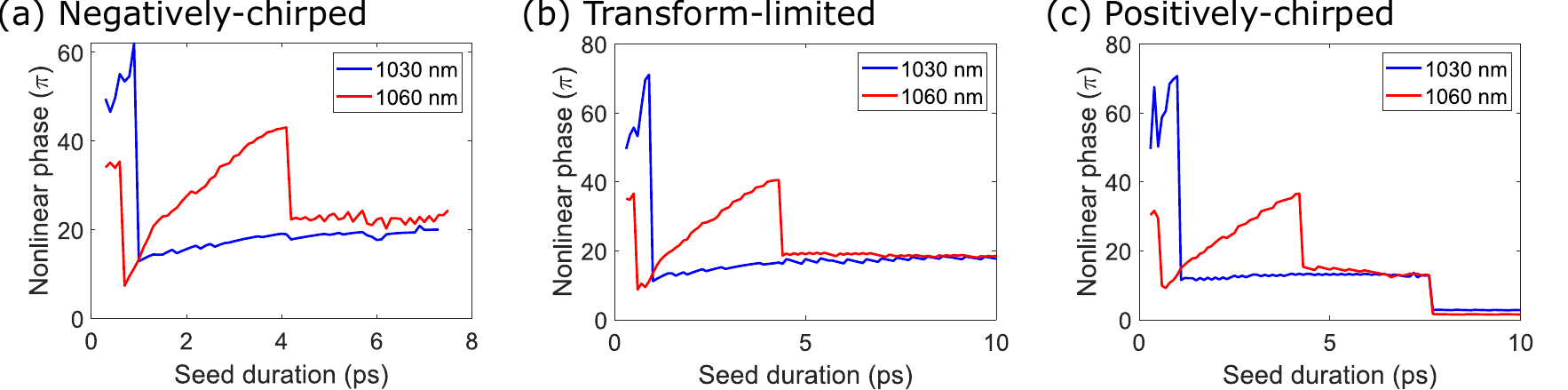}
\caption{Accumulated nonlinear phase for different chirp conditions with \ce{Yb} gain.}
\label{fig:Nonlinear_phase_Yb}
\end{figure}
\clearpage

\clearpage
\section{Fiber compressor by partial parabolic evolution}\label{sec:Fiber compressor by partial parabolic evolution}
\begin{figure}[!ht]
\centering
\includegraphics[width=\linewidth]{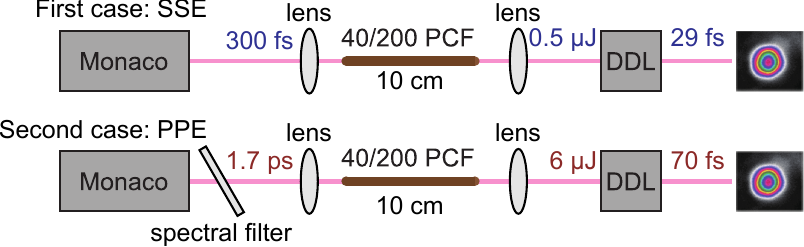}
\caption{Experimental setup. DDL: dispersion delay line, implemented with a Treacy dechirper.}
\label{fig:PPcompressor_setup}
\end{figure}

Here, we test nonlinear temporal pulse compression by exploiting nonlinear evolutions we investigated in the article with a \SIadj{10}{\cm}-long ``passive'' photonic crystal fiber [PCF (40/200 DC-Yb) from NKT Photonics; a gain fiber operated without pumping]. Investigation in a passive fiber not only removes the effect of gain bandwidth but also makes the process independent of the seed wavelength. A commercial fiber amplifier (Coherent Monaco) provides pulse energies up to \SI{60}{\micro\joule} with tunable pulse durations spanning \SI{300}{\femto\second} to \SI{10}{\pico\second}. Two representative cases are considered (Fig.~\ref{fig:PPcompressor_setup}): (1) the self-similar evolution (SSE) and (2) the partial parabolic evolution (PPE).

Similar concepts of fiber-based pulse compression were explored in prior works. Compression based on SSE produced \SIadj{0.6}{\micro\joule} and \SIadj{33}{\fs} pulses after a large-mode-area microstructured passive fiber \cite{Suedmeyer2003}. SPMA regime was studied in a long passive single-mode fiber to produce \SIadj{450}{\fs} pulses beginning at \SI{5.4}{\ps}, but the energy is very low, at only \SI{1}{\nano\joule} \cite{Nikolaus1983}. Higher \SIadj{5.2}{\micro\joule} energy was later achieved at \SI{98}{\fs} with a rod-type amplifier \cite{Saraceno2011}, but their use of \SIadj{1030}{\nm} seed lies in the narrowband high-gain region of \ce{Yb}, limiting the performance within the SPMA regime [see Fig.~3(a) in the article]. Further improvement of the performance was not possible without the complete investigation of nonlinear amplification introduced in the main article.

\begin{figure}[!ht]
\centering
\includegraphics[width=1\linewidth]{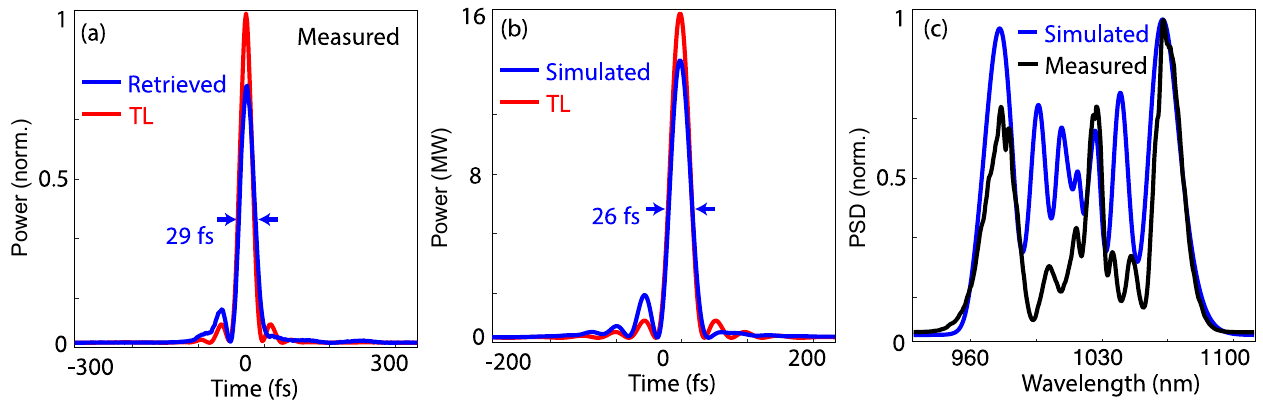}
\caption{Experimental results for the SSE-based fiber compressor. (a) FROG-retrieved pulse and (b) simulated pulse after dechirping (blue), compared with their transform-limited (TL) counterparts (red). (c) Measured (black) and simulated (blue) spectra. PSD: power spectral density.}
\label{fig:ppc29fs}
\end{figure}

In the first case, we compressed pulses with SSE by launching \SIadj{300}{\fs} pulses from the fiber amplifier. Pulses were compressed to \SIadj{29}{\fs} duration [Fig.~\ref{fig:ppc29fs}(a)] with an \SIadj{0.5}{\micro\joule} output energy. This corresponds to a peak power exceeding \SI{10}{\mega\watt} [Fig.~\ref{fig:ppc29fs}(b)]. The spectrum was broadened \num{10} times, reaching \SIadj{\sim100}{\nm} bandwidth. Our numerical simulation shows a good agreement with the experimental result [Fig.~\ref{fig:ppc29fs}(b)]. It is worth noting that it exhibits a distinct lobe on the short-wavelength side around \SI{990}{\nm}. This spectral lobe carries a pulse energy exceeding \SI{50}{\nano\joule} and may be useful for two-photon excitation of red fluorescent proteins \cite{Heikal2000}, as well as recent green fluorescent protein–based indicators such as jEDI-2P \cite{Liu2022}.

The ultimate limitation arises primarily from self-focusing. For short input pulses, the critical power can be readily reached when operating in the multi-microjoule energy regime. As predicted by the partial parabolic amplification theory, excessively-stretched pulses with durations exceeding \SI{\sim2}{\ps} exhibit reduced compressibility (Fig.~1 in the article). Therefore, next, the input pulse duration is stretched to approximately \SI{1.7}{\pico\second} for the demonstration of an optimal PPE.

Although partial parabolic evolution does not require the input pulse to exhibit a strictly parabolic temporal profile after stretching, it does require the pulse to be temporally smooth. For relatively broadband spectra of an input pulse, as in this case (\SI{\sim10}{\nano\meter}), temporal stretching to the picosecond regime by dispersion causes the temporal intensity profile to increasingly resemble the spectral profile. As a consequence, spectral ikrregularities are mapped into temporal pedestals after pre-chirping, thereby degrading the nonlinear evolution and the resulting compressibility of the output pulse.

\begin{figure}[!ht]
\centering
\includegraphics[width=0.7\linewidth]{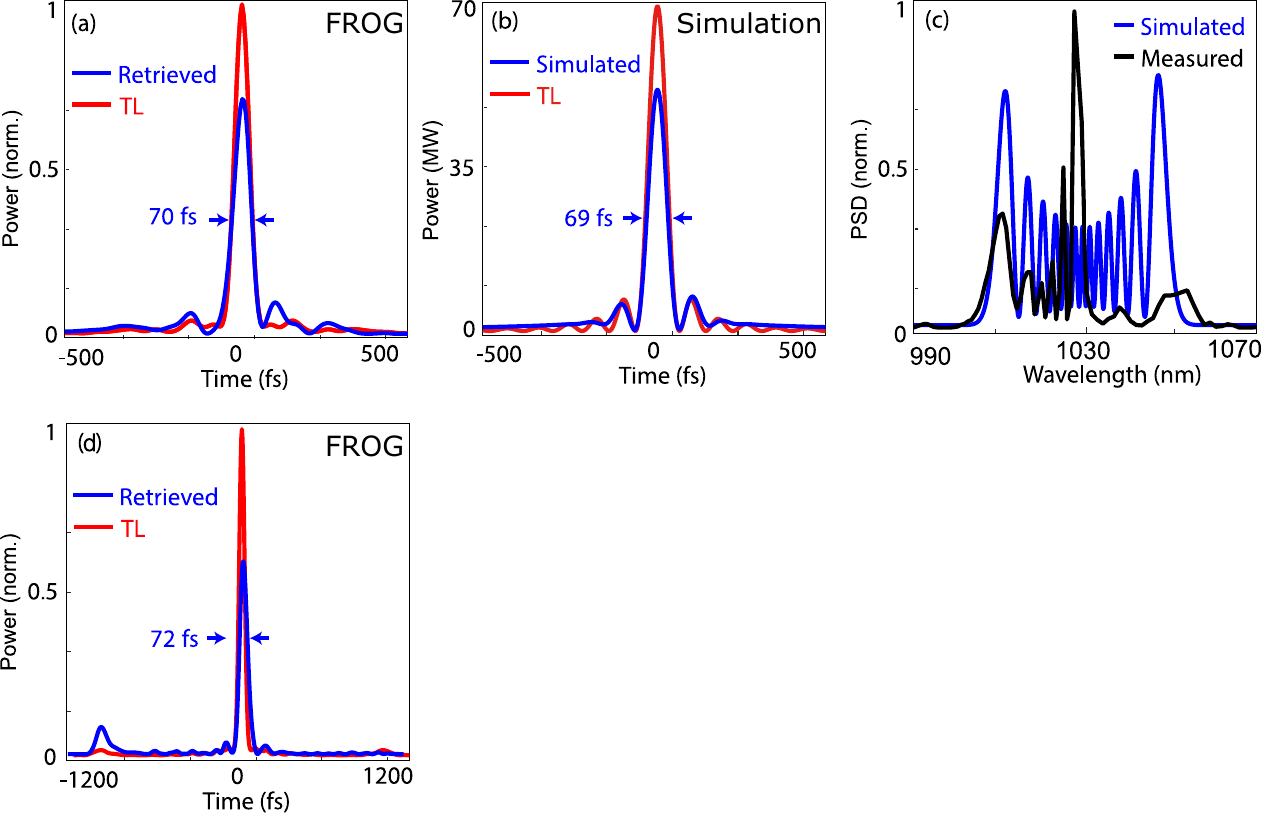}
\caption{Experimental results for the PPE-based fiber compressor. (a) FROG-retrieved pulse and (b) simulated pulse after dechirping (blue), compared with their transform-limited (TL) counterparts (red). (c) Measured (black) and simulated (blue) spectra. (d) FROG measurement with same \SIadj{1.7}{\ps} input pulse duration but without initial spectral filtering.}
\label{fig:ppc70fs}
\end{figure}

To avoid this issue, we inserted a \SIadj{4}{\nano\meter}-bandwidth bandpass filter centered at \SI{1030}{\nm} after the amplifier. This strong filtering was used for the demonstration of this technique. In practice, the bandwidth is subject to optimization to minimize the extra loss. Aggressive spectral filtering removes the fine spectral structure to yield a smooth Gaussian-like spectrum, whose temporal stretching yields a Gaussian-like temporal profile. We adjusted the pulse duration to \SI{6}{\pico\second} before filtering (by controlling the Monaco's internal dechirper stage), which led to approximately \SIadj{1.7}{\ps} duration after the spectral filter, to target the PPE regime. We were able to generate \SIadj{70}{\fs} pulses [Fig.~\ref{fig:ppc70fs}(a)], in close agreement with the numerical simulation [Fig.~\ref{fig:ppc70fs}(b, c)]. For comparison, Fig.~\ref{fig:ppc70fs}(d) shows the result obtained without spectral filtering, but directly stretching the pulses to \SI{1.7}{\ps} . Although compression to \SI{\sim70}{\femto\second} was still accomplished with the same output energy, we observed a pronounced uncompressed secondary pulse, leading to a reduced Strehl ratio of approximately \num{0.6}.

The required spectral filtering introduces an extra energy loss, which may limit the applicability of this approach if the spectrum of the source of pulses does not have a clean and smooth bell shape. However, for solid-state laser sources with intrinsically smooth spectra, or for deliberately-designed narrowband laser systems with pulse durations on the order of \SI{2}{\pico\second}, this method offers an efficient route to generating ultrashort sub-\SIadj{100}{\fs} pulses at the \SIadj{\sim10}{\micro\joule} energy level. 

\clearpage

\pagebreak
\section{Effect of chirp sign on the nonlinear amplification}\label{sec:Effect_of_chirp_sign_on_the_nonlinear_amplification}
\begin{figure}[!ht]
\centering
\includegraphics[width=0.7\linewidth]{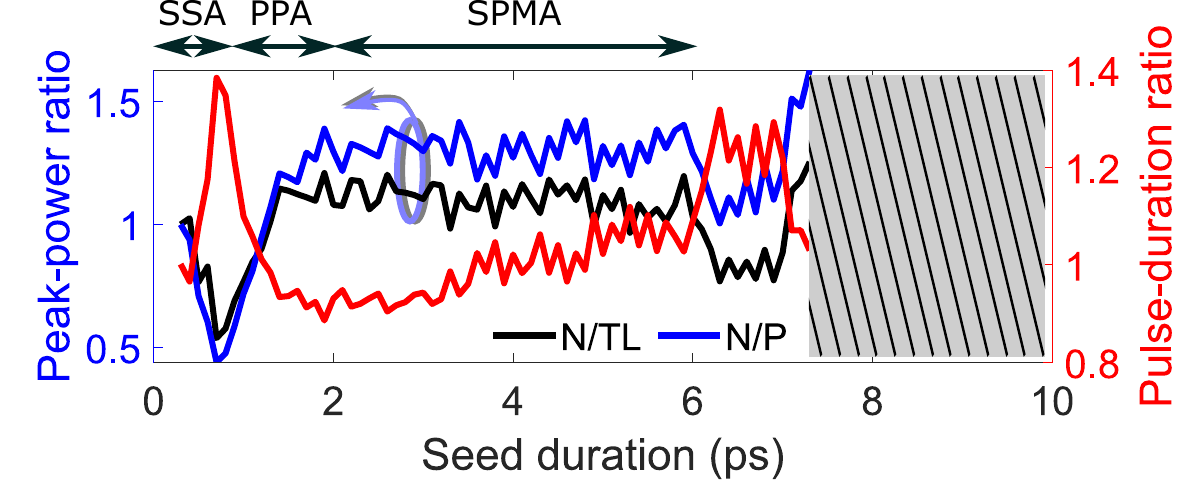}
\caption{Ratios of peak powers and durations for negatively-chirped optimization results over positively-chirped (blue; N/P) ones, in a spectrally-flat gain (Fig.~1 in the article). Additionally, peak-power ratio over transform-limited (black; N/TL) ones is shown for comparison.}
\label{fig:improvement_Gaussian}
\end{figure}

The independence of amplification from the chirp, discussed in the article, does not contradict prior results of pre-chirped amplification, which show better performance using a negatively-chirped seed than a positively-chirped seed pulse \cite{Chen2012,Zhao2014,Liu2016b,Song2017,Luo2018,Chang2019,Zhang2020a,Zhang2021b}. Our optimization in a spectrally-flat gain shows that a negatively-chirped seed can deliver approximately \SI{20}{\percent} higher peak power and \SI{8}{\percent} shorter duration of the compressed pulse than its positively-chirped counterpart around the optimal partial parabolic regime, but no improvement in duration in the SPMA regime (Fig.~\ref{fig:improvement_Gaussian}). In addition, there is no improvement in the self-similar regime due to strong dynamical parabolic shaping. \textbf{These slight improvements from a negative chirp result from an equivalently-shorter fiber}, because the pulse becomes transform-limited soon later in the propagation due to the compensation from SPM's positive chirp. Partial parabolic and SPM amplifications, which do not rely on strong parabolic shaping throughout the entire pulse, perform better with a shorter fiber. This argument is further supported by a comparison between pulses seeded with negative chirp and those with transform-limited profiles. As shown by the black line in (Fig.~\ref{fig:improvement_Gaussian}), the observed improvements diminish when benchmarked against the transform-limited counterpart. Careful examination into pre-chirped amplification shows that pre-chirping suppresses the pedestal of the compressed pulse but does not generate significantly more bandwidth \cite{Chen2012}, consistent with our observations. Prior works on pre-chirped amplification with a modest amount of negative chirp -- typically confined to the sub-picosecond regime -- lead to improved performance. However, such demonstrations should be regarded as incomplete realizations of the gain-managed nonlinear attractor due to a small amount of pre-chirp. The GMN amplification (GMNA) regime, which extends the self-similar framework, is fundamentally enabled by a dynamically varying gain profile that departs from the flat-gain condition. In earlier studies, the use of counter-pumping and short gain fibers limited the evolution of the pulse, preventing it from fully converging to the nonlinear attractor. As a result, the output remained sensitive to the initial chirp and did not exhibit the complete characteristics of GMNA evolution. It is worth noting that the incompleteness here refers to the lower achievable performance (energy and peak power) in the fiber. Typical GMNA can produce \SIadj{50}{\nano\joule} pulses from a \SIadj{6}{\micro\m}-core fiber \cite{Sidorenko2019}; those from a larger-core fiber should follow the linear scaling relation of area, \ie nonlinearity, from this energy value. However, in practice, GMNA faces challenges in scaling beyond \SI{1}{\micro\joule} \cite{Sidorenko2020} because a fiber with a much-larger core is mostly rod-type and needs to be physically straight, which makes complete GMN evolution in a \SIrangeadj{2}{4}{\m}-long rod-type fiber unrealistic. Hence, \textbf{pre-chirped amplification that undergoes GMN-like evolution within a shorter fiber was, in fact, an attractive alternative to a perfect GMNA in scaling beyond \SI{1}{\micro\joule} \cite{Zhao2014,Liu2015,Liu2016b,Zhang2020a,Zhang2021b} -- before GMPPA is introduced in this work}.

\begin{figure}[!ht]
\centering
\includegraphics[width=0.7\linewidth]{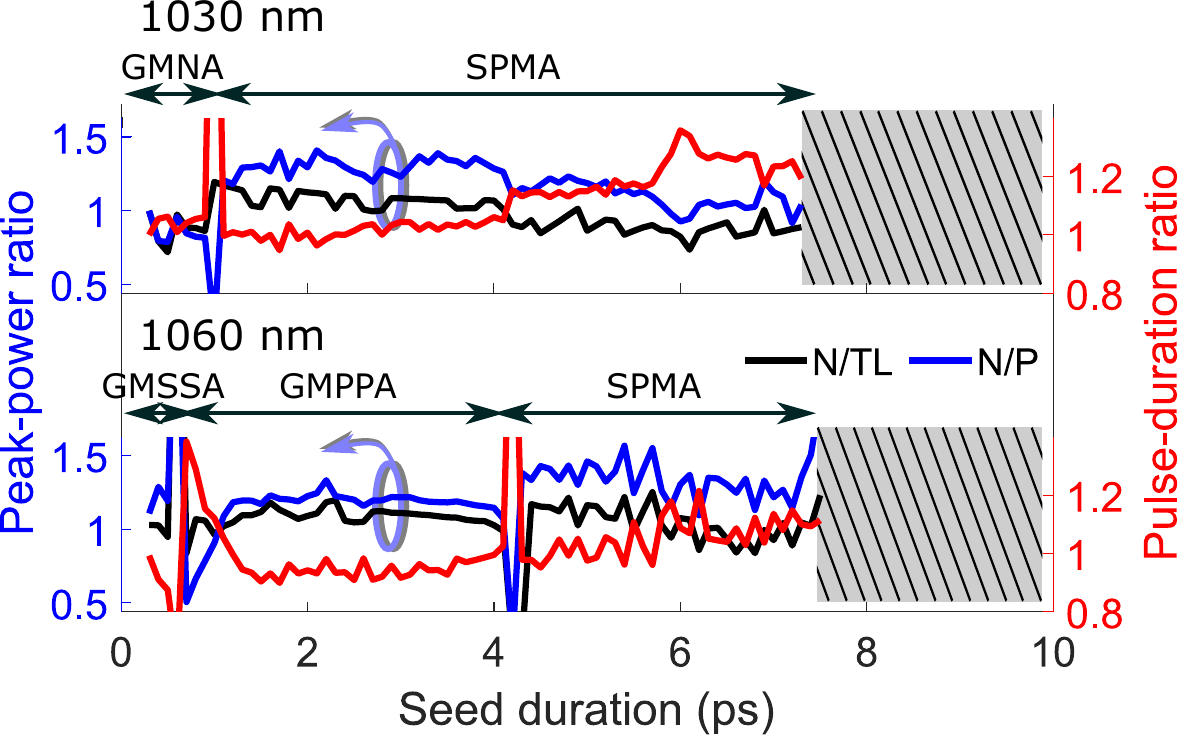}
\caption{Ratios of peak powers and durations for negatively-chirped optimization results over positively-chirped (blue; N/P) ones, in a \ce{Yb}-doped fiber amplifier (Fig.~3 in the article). Additionally, peak-power ratio over transform-limited (black; N/TL) ones is shown for comparison.}
\label{fig:improvement_Yb}
\end{figure}

Additionally, we observe the same improvement trend as in flat-gain scenarios with a negatively-chirped seed over its positively-chirped counterpart, which gives \SI{20}{\percent} higher peak power and \SI{7}{\percent} shorter duration of the compressed pulse in the GMSSA and GMPPA regimes, as well as \SI{30}{\percent} higher peak power but no improvement in duration in the SPMA regime; no improvement is observed in the GMNA regime (Fig.~\ref{fig:improvement_Yb}).
\clearpage

\pagebreak
\section{Transition between GMPPA and SPMA regimes}\label{sec:Transition_between_GMPPA_and_SPMA_regimes}
In this section, we explain in detail the sharp transition occurring between GMPPA and SPMA regimes in Fig.~3(b) in the article. Gain management plays an important role in creating such a sharp transition, which would, otherwise, become a smooth transition as in a flat gain scenario (Fig.~1).

\begin{figure}[!ht]
\centering
\includegraphics[width=0.7\linewidth]{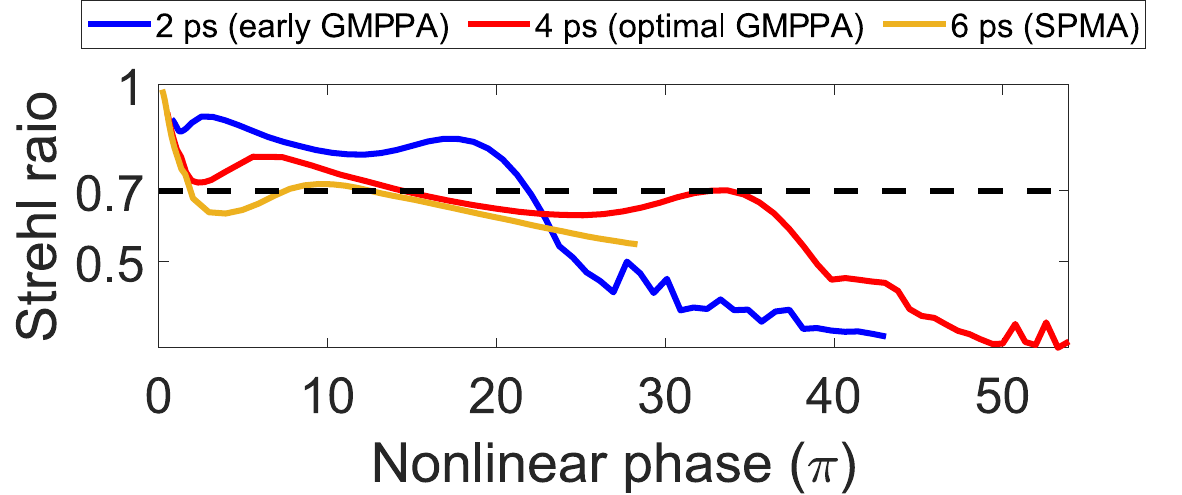}
\caption{Variations of Strehl ratio with different nonlinear phase of the \SIadj{1060}{\nm} process in \ce{Yb} in Fig.~3(b). We use ``nonlinear phase'' here as it broadly involves the effects from pump power, seed energy, and seed duration. Nonlinear phase is approximated from $\phi_{\text{NL}}=\gamma\int P_{\text{peak}}\diff z$, where $\gamma$ is the nonlinear coefficient at the pulse's center frequency and $P_{\text{peak}}$ is the peak power. Here, we run simulations with varying pump powers, where nonlinear phase represents monotonically the pump power, with higher pump power corresponding to higher nonlinear phase.}
\label{fig:GMPPA_SPMA_transition}
\end{figure}

As explained in the article, the GMPPA regime exhibits TOD compensation from the gain management when the spectrum reaches the short-wavelength high-gain region in \ce{Yb}. Due to this process, the TOD compensation is diminished when the pump power drops, leading to a reduced Strehl ratio. To illustrate the transition, we consider three durations: \SI{2}{ps}, \SI{4}{ps}, and \SI{6}{ps} representing early-stage GMPPA, optimal GMPPA, and SPMA regimes, respectively [where detailed parameters are taken from Fig.~3(b) in the article]. Since we constrain the minimum Strehl ratio to \num{0.7}, in the early-stage GMPPA (blue line in Fig.~\ref{fig:GMPPA_SPMA_transition}), the pulse is amplified to an extent such that even the gain-managed third-order-dispersion (TOD) compensation starts to fail (reducing Strehl ratio at around \num{22\pi} in the blue line). As of the optimal GMPPA (red line in Fig.~\ref{fig:GMPPA_SPMA_transition}), the process exhibits a lower Strehl ratios throughout the whole parameter space of pump power (represented by nonlinear phase in Fig.~\ref{fig:GMPPA_SPMA_transition}) due to an increased seed duration. The optimum occurs when gain-managed TOD compensation balances mismatched TOD from the nonlinearity and the Treacy dechirper. As we can see in this process, a decrease of pump power or an increase of seed duration will lead to a further reduction in Strehl ratio, resulting in a failure to find any good Strehl ratio (\num{>0.7}) in the regime of high nonlinear phase. Therefore, the process is forced to reduce significantly the nonlinear phase to maintain a good Strehl ratio, \eg to those around \num{10\pi} in Fig.~\ref{fig:GMPPA_SPMA_transition}. The pulse may then re-adjust itself for an optimal state around a lower nonlinear phase, thus evolving into the SPMA regime (orange line in Fig.~\ref{fig:GMPPA_SPMA_transition}).

It is worth noting that variation of the Strehl ratio with increased nonlinear phase result not only from gain-managed TOD compensation but also from the balance between quadratic and cubic phases in the pulse. A decent amount of cubic phase can help shape the pulse with quadratic phase, increasing the peak power. Therefore, a practical (imperfect) dechirping does not aim to remove all quadratic phase, leaving only cubic phase, but aim toward a phase with balanced quadratic and cubic phases such that the peak power is the highest. Because the ratio of quadratic-to-cubic phases of the pulse and the Treacy dechirper are not matched, Strehl ratio can fluctuate, rather than monotonically varying, with an increased nonlinear phase to compensate for in the process. This explains the fluctuation of Strehl ratio in the SPMA regime (orange line in Fig.~\ref{fig:GMPPA_SPMA_transition}) where gain management does not take effect. It also explains the initial fluctuations at low nonlinear phase in the GMPPA regime (blue and red lines in Fig.~\ref{fig:GMPPA_SPMA_transition}).

\clearpage

\pagebreak
\section{\num{30}-\si{\micro\m} large-mode-area fiber results and peak-power test}\label{sec:Peak_power_test_for_the_shortest_compressed_pulses}
\subsection{\num{30}-\si{\micro\m} large-mode-area-fiber results}
\begin{figure}[!ht]
\centering
\includegraphics[width=0.8\linewidth]{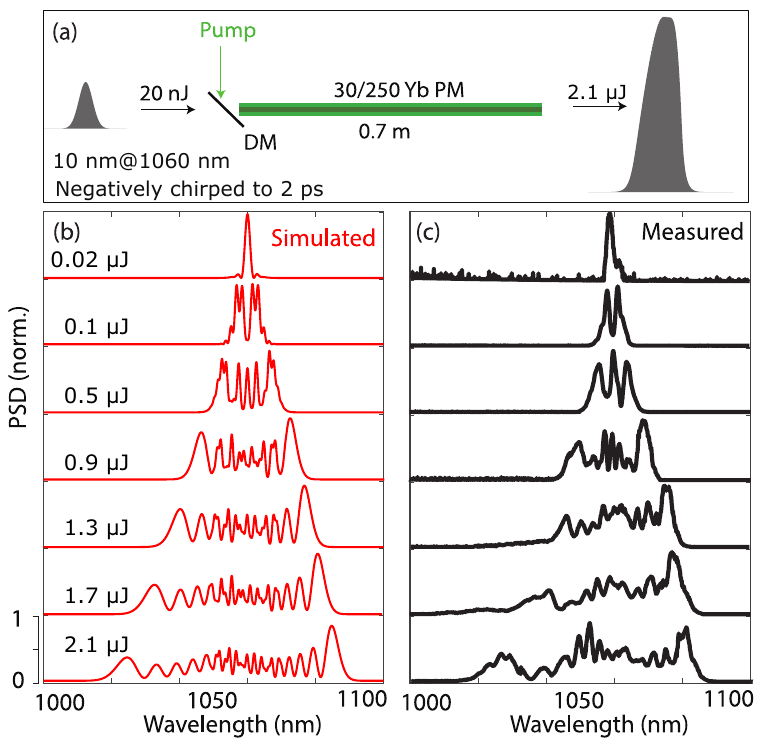}
\caption{Experimental validation of the GMPPA regime with a \SIadj{30}{\micro\m}-core fiber. (a) Schematics of the experimental setup. DM: dichroic mirror. (b) Simulated and (b) measured output spectra for pulse energies ranging from \num{0.02} to \SI{2.1}{\micro\joule}.}
\label{fig:Measurements2}
\end{figure}

\begin{figure}[!ht]
\centering
\includegraphics[width=0.7\linewidth]{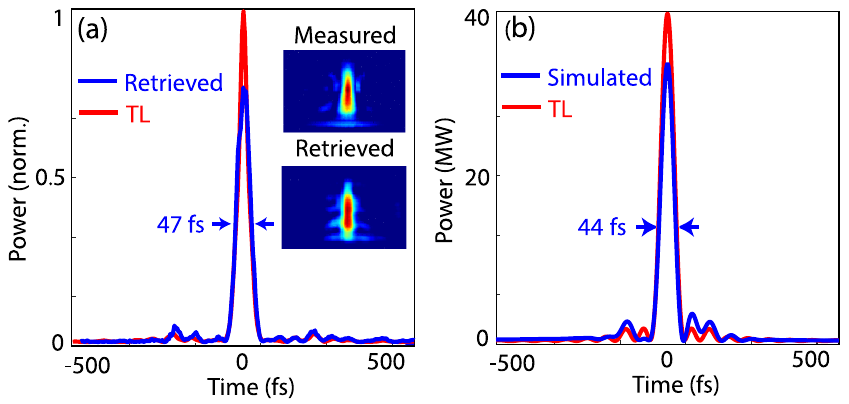}
\caption{Temporal measurements. (a) FROG measurement and (b) simulated dechirped pulse (blue), compared with their transform-limited (TL) counterparts (red).}
\label{fig:FROG_supplement}
\end{figure}

Guided by the primary experimental validations, we performed supplementary measurements using a \SIadj{70}{\cm}-long \num{30}/\num{250} \ce{Yb}-doped fiber (Coherent, PLMA-YDF-30/250-UF). The pump absorption is \SI{4}{\decibel/\m} at \SI{915}{\nm}, and fiber core diameter is \SI{30}{\micro\m}. The energy of the seed pulse remained at \SI{20}{\nano\joule}, but it was pre-chirped to approximately \SI{1.6}{\ps}. The experimental (black) and simulated (red) spectra for different pulse energies are shown in Fig.~\ref{fig:Measurements2}. These spectra exhibit consistent broadening and modulation patterns, with simulations closely matching experimental data across the entire energy range. Notably, this setup achieved about the same pulse energy of \SI{2.1}{\micro\joule} with comparable duration (\SI{47}{\fs}, Fig.~\ref{fig:FROG_supplement}) and spectral broadening as the \SIadj{25}{\micro\m} fiber in the article. Here, we are limited by the available pump power for the gain fiber. Otherwise, we expect to obtain higher energy if the pulse is stretched to \SI{3}{\ps} as in the article. 

\pagebreak
\subsection{Peak-power test}
\begin{figure}[!ht]
\centering
\includegraphics[width=0.8\linewidth]{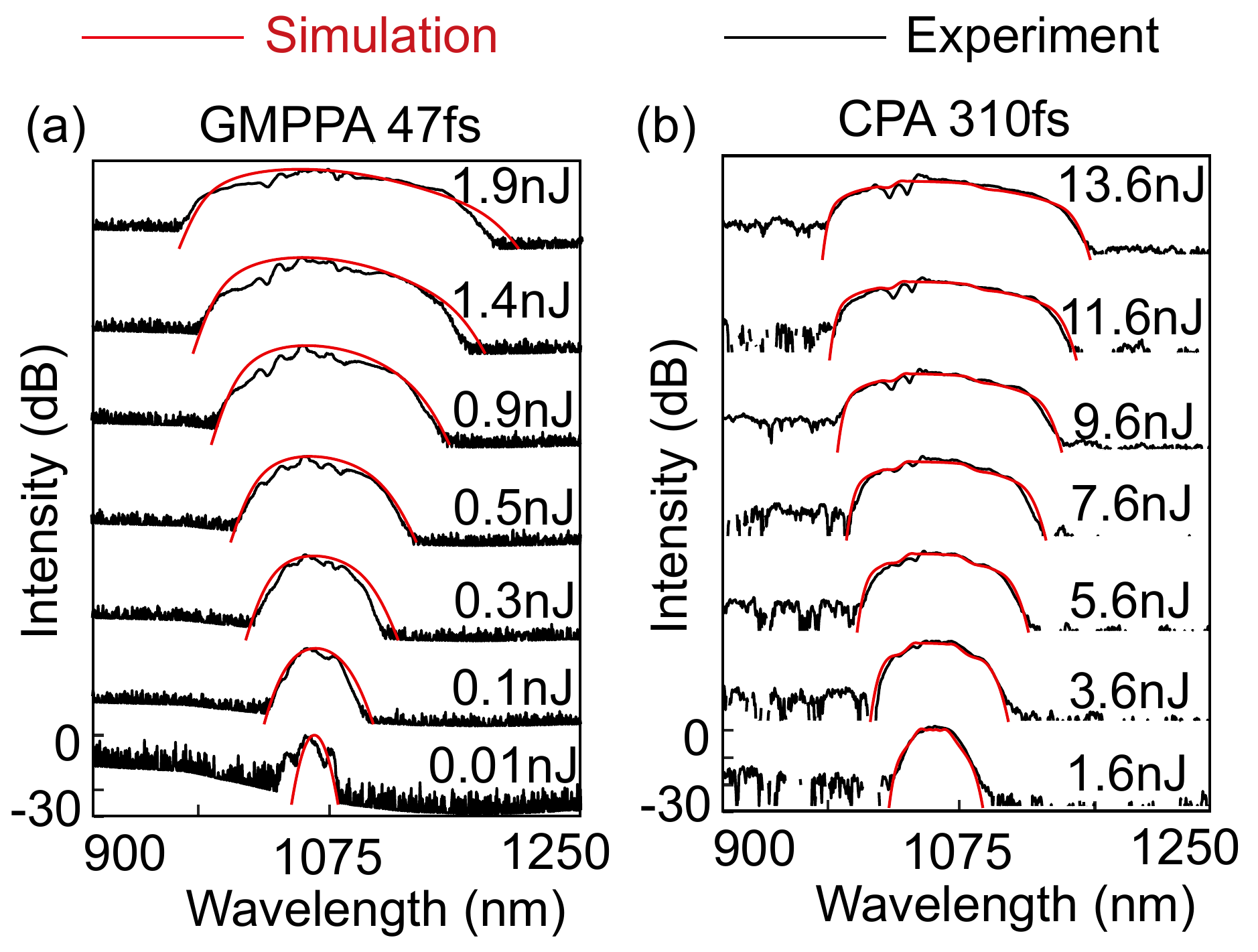}
\caption{Experimental (black) and simulated (red) spectra after \SIadj{1}{\m}-long PM980 fiber measured with the indicated pulse energies. (a) Results with the \SIadj{47}{\fs} pulses from the GMPPA. (b) Results with the \SIadj{310}{\fs} pulses from a home-built fiber CPA.}
\label{fig:fs3}
\end{figure}

We evaluate the quality of the compressed pulses based on nonlinear pulse evolution. A portion of the dechirped pulses was coupled into a \SIadj{1}{\m} segment of polarization-maintaining fiber (PM980), and the spectrum at the fiber output end was measured as a function of input pulse energy. The excellent agreement between the measured and simulated spectra, as depicted in Fig.~\ref{fig:fs3}(a), serves as a validation of the \SIadj{30}{\MW} peak power achieved with the \SIadj{2.1}{\micro\joule} and \SIadj{47}{\fs} pulses. Additionally, we conducted a peak-power test using our home-built fiber chirped-pulse amplifier (CPA) system, which generates pulses with a \SIadj{310}{\fs} duration. For this system, we coupled pulse energies ranging from \num{1.6} to \SI{13.6}{\nano\joule} into the PM980 fiber, yielding the broadened spectra shown in Fig.~\ref{fig:fs3}(b). These results exhibit comparable spectral broadening but at significantly higher pulse energies -- \eg \SI{1.9}{\nano\joule} for \SIadj{47}{\fs} pulses versus \SI{13.6}{\nano\joule} for \SIadj{310}{\fs} pulses -- and highlight the benefit of the substantially-shorter pulses produced by GMPPA.

\clearpage

\pagebreak
\section{Effect of seed pulse duration}\label{sec:seed_duration}
\begin{figure}[!ht]
\centering
\includegraphics[width=0.8\linewidth]{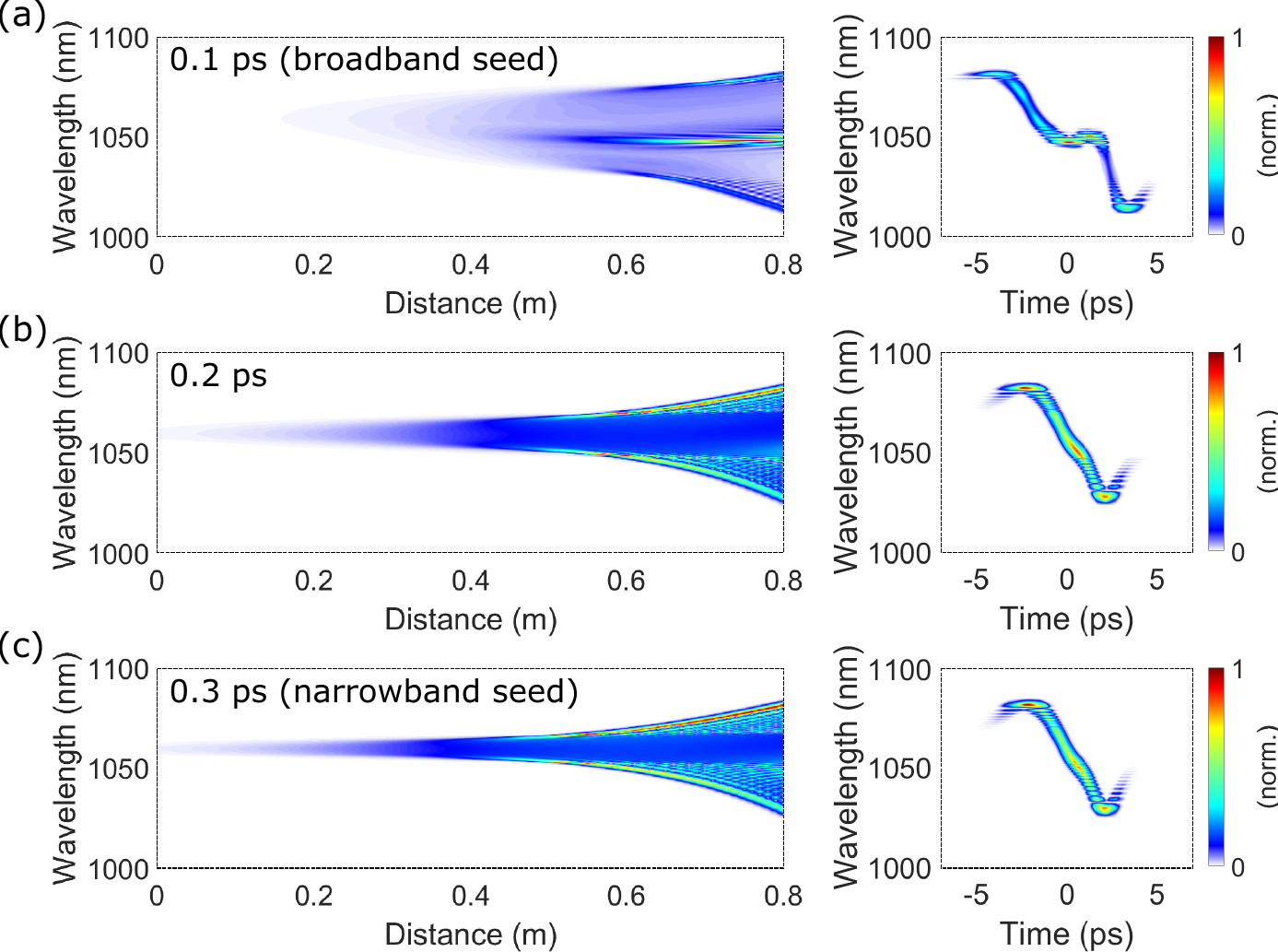}
\caption{Bandwidth limit of the seed to evolve in the GMPPA regime. Spectral evolutions and spectrograms of the amplified pulses are shown for \SIadj{1060}{\nm} seeds with (a) \SIadj{0.1}{\ps}, (b) \SIadj{0.2}{\ps}, and (c) \SIadj{0.3}{\ps} transform-limited durations. Seeds are positively chirped to \SI{4}{\ps} for GMPPA.}
\label{fig:seed_bandwidth_limit}
\end{figure}

In the article, we fixed the minimum seed duration to \SI{0.3}{\ps}. However, the bandwidth of the seed can significantly affect the nonlinear amplification (Fig.~\ref{fig:seed_bandwidth_limit}). Transition into the GMPPA regime requires that the pulse initially evolves under the dynamics of the PPA regime, with gain management subsequently acting as a reshaping mechanism during later stages of propagation. If the pulse reaches the \SIadj{1030}{\nm} high-gain region in the early stage, differential gain in two spectral regions induces distinct nonlinear phase modulations, leading to different chirps for the high- and low-gain regions [Fig.~\ref{fig:seed_bandwidth_limit}(a)]. Such phase cannot be compensated by a Treacy dechirper. On the other hand, pulses with a transform-limited duration longer than \SI{0.2}{\ps} can successfully evolve into the GMPPA regime [Figs.~\ref{fig:seed_bandwidth_limit}(b) and (c)].

\pagebreak
\section{Effect of seed pulse wavelength}\label{sec:seed_wavelength}

\subsection{Opposite direction of broadening with the seed at \SI{1000}{\nm}}\label{subsec:Opposite broadening for 1000nm seeded case}
\begin{figure}[!ht]
\centering
\includegraphics[width=0.7\linewidth]{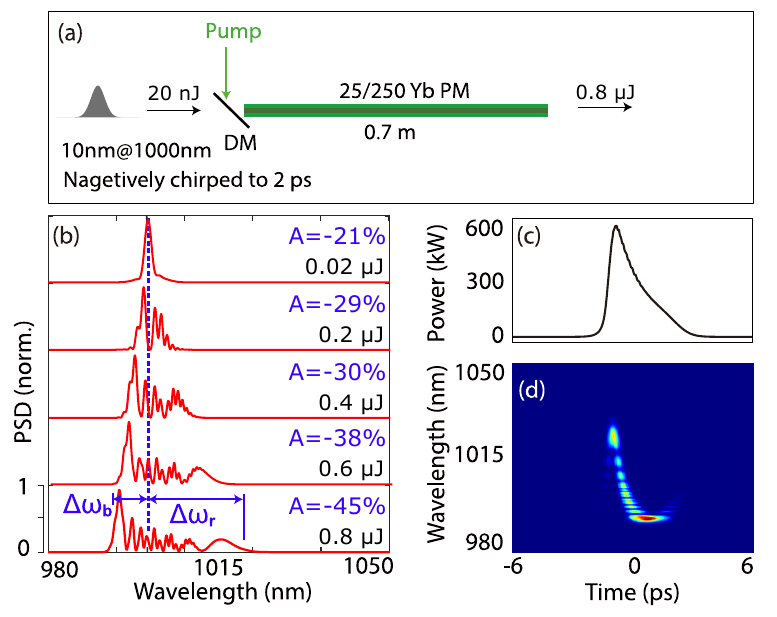}
\caption{Simulation of the \SIadj{1000}{\nm} seed case. (a) Schematic of the simulated system. DM: dichroic mirror. (b) Simulated output spectra for pulse energies ranging from \num{0.02} to \SI{0.8}{\micro\joule}. $A$ is the asymmetry parameter defined in the article. (c) Temporal profile of the pulse before dechirping and (d) its spectrogram.}
\label{fig:1000nm}
\end{figure}

In this section, we numerically demonstrate the spectral broadening with a seed at \SI{1000}{\nm} (Fig.~\ref{fig:1000nm}). This aims to showcase the gain management that generates the opposite sign of TOD, compared to the GMPPA discussed in the article. It facilitates a deeper understanding of gain management and its relation with the produced TOD, which is crucial for TOD compensation with a dechirper.

A \SIadj{20}{\nano\joule} seed is launched into a \ce{Yb}-doped fiber, with the same parameters as those in the article [Fig.~\ref{fig:1000nm}(a)]. It is negatively chirped to \SI{2}{\ps} for similar evolution to the GMPPA. Due to the high-gain region around \SI{1030}{\nm}, the pulse preferentially broadens toward the long-wavelength side [Fig.~\ref{fig:1000nm}(b)]. This creates a pronounced sharply-rising leading edge [Fig.~\ref{fig:1000nm}(c)], also shown by the spectrogram [Fig.~\ref{fig:1000nm}(d)], where the long-wavelength components all stay within the leading edge. TOD generated by this pulse shape has the same sign as a Treacy grating dechirper, so the pulse cannot be fully dechirped with it. Dechirping such a pulse requires a dechirper with a negative sign of TOD, such as a grism-based or prism-based pulse dechirper.

\subsection{Optimal nonlinear amplification regimes seeded at \SI{1050}{\nm}}\label{subsec:Optimal_nonlinear_amplification_regimes_seeded_at_1050nm}
\begin{figure}[!ht]
\centering
\includegraphics[width=\linewidth]{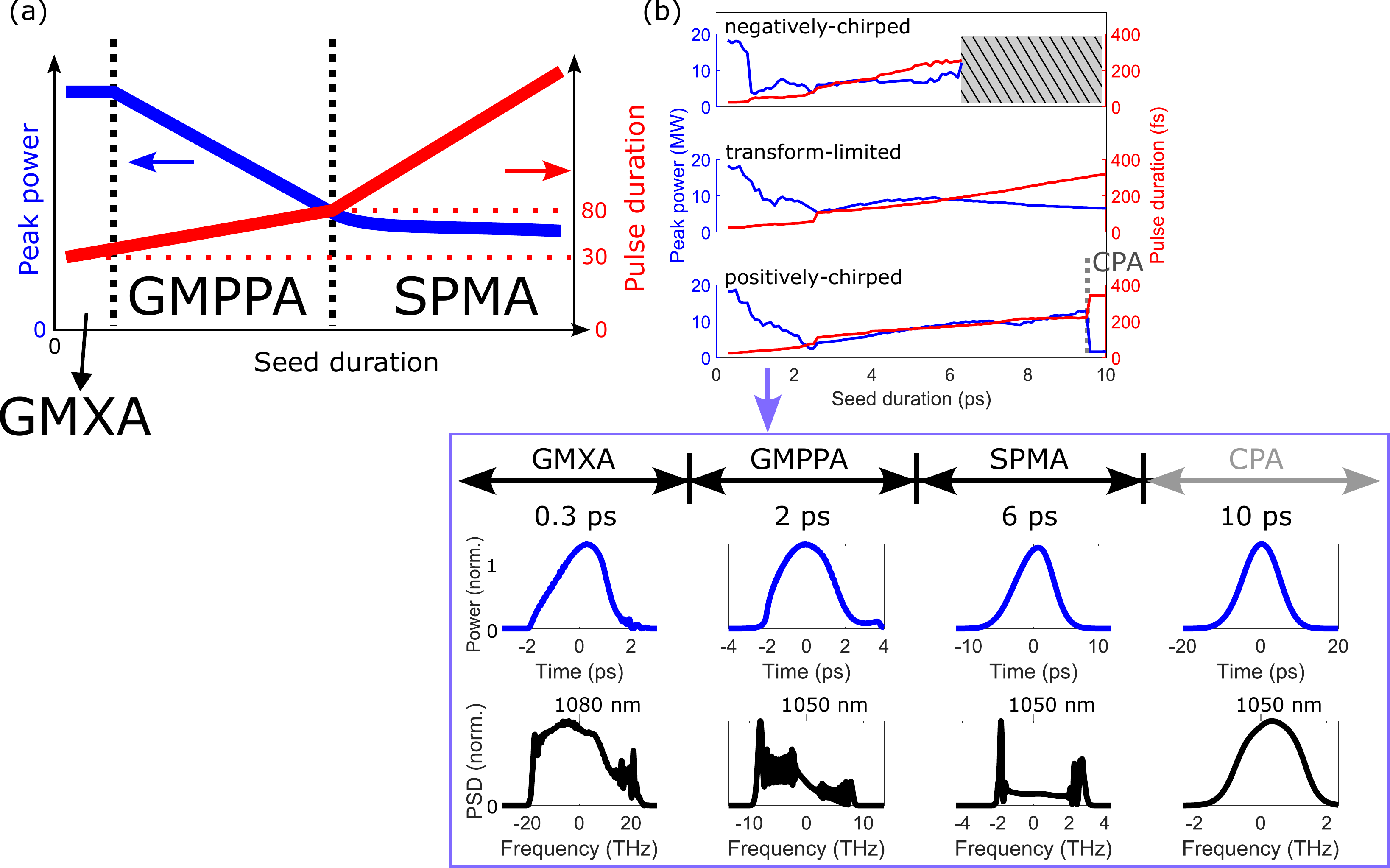}
\caption{Optimal nonlinear amplification regimes at different seed durations in \SIadj{25}{\micro\m}-core \ce{Yb}-doped fiber for a \SIadj{1050}{\nm} seed. (a) Schematic of the optimal nonlinear amplification regimes, displayed as peak power and duration of the dechirped amplified pulse at different seed durations. GMXA represents the intermediate regime between GMNA and GMSSA. (b) Optimization results for negatively-chirped, transform-limited, and positively-chirped seeds. Below are the temporal and spectral profiles of the amplified pulses for regimes other than GMNA, for the indicated durations of a positively-chirped seed. Spectra are plotted with respect to frequency relative to the center frequency.}
\label{fig:regime_Yb_1050nm}
\end{figure}

Here, we show the optimization results with a \SIadj{1050}{\nm} seed (Fig.~\ref{fig:regime_Yb_1050nm}). As \SI{1050}{\nm} is found to be the optimal wavelength for linear CPA \cite{Schimpf2010}, it is natural to ask whether this is also true for nonlinear amplification. The boundary of the high-gain region for Yb is at \SI{1050}{\nm}, so the performance may be distinct from that at \SI{1030}{\nm} and \SI{1060}{\nm}.

With a short seed pulse, amplification is an intermediate case between GMNA and GMSSA. The \SIadj{1050}{\nm} seed initially experiences strong amplification due to being near the high-gain region. Similar to GMNA, later gain saturation leads to absorption of its short-wavelength components and gain-managed TOD compensation. The pulse can be dechirped to \SI{25}{\fs} and approximately \SIadj{20}{\MW} peak power, similar to best results for GMNA and GMSSA (Fig.~3 in the article).

Due to the effectively-broad gain spectrum for \SIadj{1050}{\nm} light, a longer seed pulse undergoes GMPPA evolution, and the spectrum exhibits features as in GMPPA seeded at \SI{1060}{\nm}. Unlike GMSSA, in which later gain-managed amplification introduces a proper amount of TOD compensation, the \SIadj{1050}{\nm} pulse experiences the high-gain amplification at the beginning of the propagation. This produces excess TOD that over-compensates the cubic phase introduced by the Treacy dechirper. The bandwidth of the pulse needs to be reduced to alleviate the excessive effect of gain management for optimal TOD compensation. Therefore, GMPPA seeded at \SI{1050}{\nm} produces pulses with lower dechirped peak power and longer pulse duration than when seeded at \SI{1060}{\nm}.

\subsection{Optimal nonlinear amplification regimes seeded at wavelengths longer than \SI{1050}{\nm}}\label{subsec:Optimal_nonlinear_amplification_regimes_seeded_at_wavelengths_longer_than_1050nm}
\begin{figure}[!ht]
\centering
\includegraphics[width=0.8\linewidth]{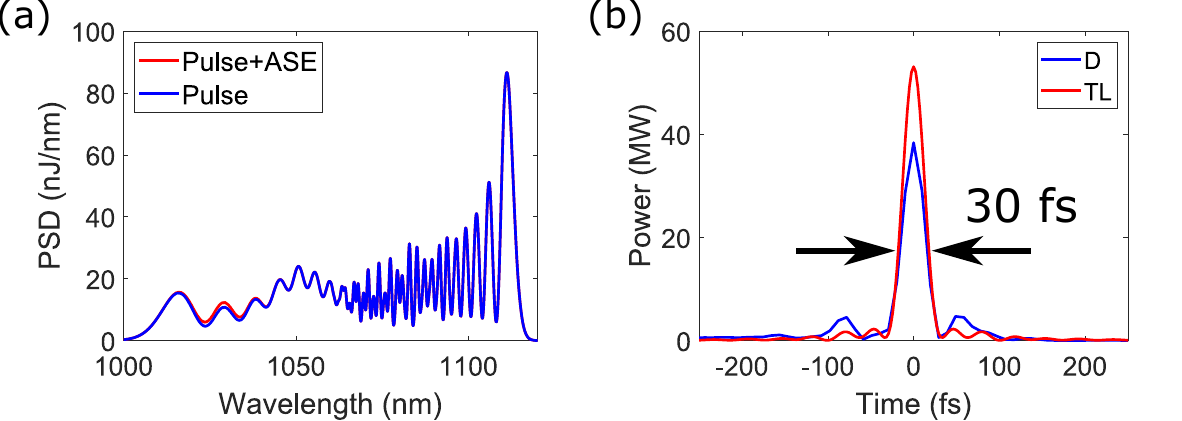}
\caption{(a) Simulated spectrum and (b) dechirped (D) and its transform-limited (TL) profiles of the \SIadj{1.9}{\micro\joule} compressed pulse seeded at \SI{1080}{\nm}. Seed has \SIadj{60}{\nano\joule} energy with \SIadj{0.2}{\ps} transform-limited duration, negatively chirped to \SI{1.6}{\ps}. Pump power is \SI{11}{\W}. Fiber length is \SI{0.7}{\m}. Strehl ratio is \num{0.73} for the dechirped pulse. ASE contributes to only \SI{1.3}{\percent} of the total power.}
\label{fig:1080nm}
\end{figure}

Throughout investigations of \ce{Yb}-doped amplifier in the article, we select a long seed wavelength of \SI{1060}{\nm} to study nonlinear amplification within the flat-gain region of \ce{Yb}. With a seed at longer wavelengths, gain management can become less effective, as the pulse requires a higher pump power to spectrally extend into the high-gain region below \SI{1050}{\nm}. This is attributed to the reduced gain of \ce{Yb} at longer wavelengths and the increased spectral broadening necessary to reach the high-gain region. Eventually, the behavior of nonlinear amplification follows the flat-gain scenarios with a longer-wavelength seed, resulting in a limited performance. In addition, the process might generate excess ASE due to a lower signal gain compared to the ASE gain at \SI{1030}{\nm}. Nevertheless, all these can be overcome by sacrificing gain and using an energetic seed pulse. Seeding at a longer wavelength can generate the broadest GMPPA spectrum, as gain management restricts the extent to short wavelengths but not to long wavelengths. With only \num{30} times amplification, pulses seeded at \SI{1080}{\nm} in the GMPPA regime can numerically achieve \SIadj{30}{\fs} duration, approaching the GMNA limit, with \SIadj{40}{\MW} peak power (Fig.~\ref{fig:1080nm}).

\clearpage

\pagebreak
\section{Re-classification of prior works}\label{sec:Reclassification_of_prior_works}
This section surveys prior studies pertinent to nonlinear amplification accompanied by substantial pulse compression, as summarized in Tables \ref{tab:categorization1}--\ref{tab:categorization3}. With the understanding of nonlinear amplification presented in the article, we think it is insightful to re-classify previous work on nonlinear amplification. Furthermore, we indicate whether each study operates within the optimal regime. Of course, the re-classification is done with perfect hindsight, and is not intended to suggest that prior works made erroneous claims at the time they were published.

It is important to re-classify pre-chirped amplification for a better understanding. Prior works targeted the self-similar amplification \cite{Zhao2014,Liu2016b,Luo2018,Zhang2020a}; however, the narrowband gain at \SI{1030}{\nm} does not support broadband pulse generation for sub-\SIadj{40}{\fs} compressed pulses. Pre-chirped amplification, which is, in fact, incomplete evolution of GMNA (a detailed discussion is in Sec.~\ref{sec:Effect_of_chirp_sign_on_the_nonlinear_amplification}) if it reaches sub-\SIadj{40}{\fs} durations with a \SIadj{\sim1030}{\nm} seed, generates up to \SIadj{51}{\MW} \cite{Liu2016b} and \SIadj{<55}{\MW} \cite{Zhang2021b} peak powers in \SIadj{40}{\micro\m}-core and \SIadj{85}{\micro\m}-core fibers, respectively. However, after linear scaling to \SIadj{25}{\micro\m} core for a fair comparison, they correspond to \SI{20}{\MW} and \SI{<4.8}{\MW}, respectively, smaller than our initial demonstration of \SI{30}{\MW} with GMPPA in the article. Compared to GMNA that requires a long fiber to operate, GMPPA can use only a short gain fiber, enabling simple scaling with commercially-available large-mode-area or rod-type fibers that are typically fabricated in short lengths. In principle, prior pre-chirped demonstrations may be revisited and scaled using GMPPA by temporally stretching the seed pulse to the picosecond regime and incorporating a long-wavelength seed pulse.

As Mamyshev oscillators push toward the boundary of nonlinear amplification, here we also include some of their early works. The amplification before the output port of a Mamyshev oscillator can be treated as a single-pass fiber amplifier whose seed wavelength is determined by the filter. We do not consider other oscillators, as they are restricted by other conditions, including multi-pulsing, or requirement of periodic condition, before getting close to the nonlinear amplification limit. For example, a typical cat-ear-shaped all-normal-dispersion oscillator \cite{Chong2006,Renninger2008} is simply an outcome of a positively-chirped pulse undergoing SPM, where the central part of the spectrum is free from spectral modulations \cite{Finot2018}. Attempts to obtain a higher-energy state in these oscillators result in multi-pulsing or breakdown of stable mode-locking.

\begin{figure}[!ht]
\centering
\includegraphics[width=0.7\linewidth]{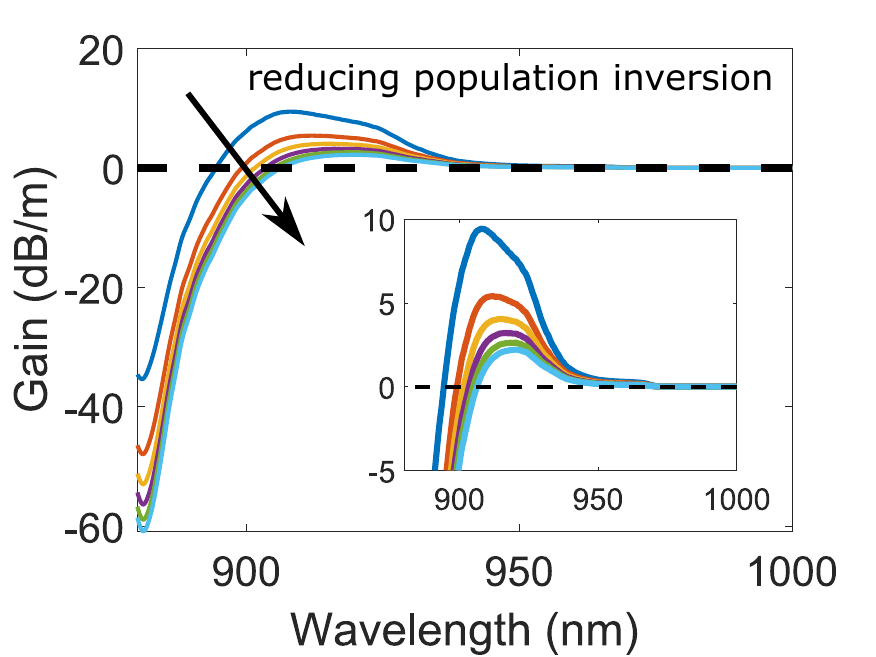}
\caption{Calculated gain spectrum of \ce{Nd}-doped fiber amplifier with reduced population inversion (or increased gain saturation). Pump is at \SI{808}{\nm}. This is generated by our publicly-shared code on GitHub, considering up to \num{12} levels in \ce{Nd}. The seed needs to be below \SI{907}{\nm} to experience loss at high saturation for GMNA or above \SI{940}{\nm} to experience short-wavelength amplification for GMSSA and GMPPA to work as expected.}
\label{fig:Nd_gain_spectrum}
\end{figure}

\clearpage 
\begin{table}[!ht]
\hspace{-10em}
\begin{tabular}{c|ccccl}
\toprule
\textbf{Prior work} & \parbox{0.18\textwidth}{\textbf{Its regime}/\\ \textbf{Optimal regime at this duration}} & \parbox{0.17\textwidth}{\textbf{Doped ion}/\\ \textbf{Core diameter}/\\ \textbf{Fiber length}} & \parbox{0.18\textwidth}{\textbf{Seed duration}/\\ \textbf{\hspace*{2em}wavelength}} & \parbox{0.267\textwidth}{\textbf{Compressed duration}/\\ \textbf{\hspace*{5em}peak power}} & \textbf{Notes} \\
\midrule
\cite{Fermann2000} & SSA/GMSSA & \ce{Yb}/\SI{6}{\micro\m}/\SI{3.6}{\m} & \SI{200}{\fs}/\SI{1060}{\nm} & \SI{68}{\fs}/\SI{80}{\kW} & First demonstration of SSA. \\[1em]
\cite{Limpert2002a} & SSA/GMSSA & \ce{Yb}/\SI{30}{\micro\m}/\SI{9}{\m} & \SI{180}{\fs}/\SI{1060}{\nm} & \SI{80}{\fs}/\SI{1.7}{\MW} & \SIadj{227}{\nano\joule}\textsuperscript{\footnotesize*} output. \\[1em]
\cite{Papadopoulos2007} & SSA/GMXA & \ce{Yb}/\SI{20}{\micro\m}/\SI{6.5}{\m} & \SI{145}{\fs}/\SI{1040}{\nm} & \SI{63}{\fs}/\SI{4.1}{\MW} & \parbox{0.36\textwidth}{Study SPM-induced TOD compensation. \SIadj{290}{\nano\joule} output.} \\[2em]
\cite{Zaouter2007} & SSA/GMXA & \ce{Yb}/\SI{25}{\micro\m}/\SI{4}{\m} & \SI{87}{\fs}/\SI{1050}{\nm} & \SI{107}{\fs}/\SI{<340}{\kW}\textsuperscript{\footnotesize*} & \parbox{0.36\textwidth}{Apply prism-based TOD compensation. \SIadj{38}{\nano\joule} output.} \\[1em]
\cite{Zaouter2008} & SPMA/GMNA & \ce{Yb}/\SI{80}{\micro\m}/\SI{85}{\cm} & \SI{330}{\fs}/\SI{1030}{\nm} & \SI{70}{\fs}/\SI{16}{\MW} & \parbox{0.36\textwidth}{\SIadj{1.25}{\micro\joule} output} \\[1em]
\cite{Schreiber2006} & \parbox{0.18\textwidth}{\color{magenta}(SPMA-CPA)/\\ \hspace*{2em}CPA} & \ce{Yb}/\SI{30}{\micro\m}\textsuperscript{\footnotesize*}/\SI{1.2}{\m} & \SI{\sim10}{\ps}/\SI{1035}{\nm} & \SI{240}{\fs}/\SI{5}{\MW} & \parbox{0.36\textwidth}{Parabolic pre-shaping due to temporal stretching of a \SIadj{240}{\fs} transform-limited pulse to \SI{\sim10}{\ps}. This work targets CPA for a parabolic seed. The \SI{10}{\ps} they use has a strong coincidence, as the SPMA process with a consistent \num{\sim30} times compression ratio can compress it back to the original \SI{240}{\fs}. At this duration, CPA regime overlaps with SPMA to some extent [Fig.~3(a) in the article]. \SIadj{1.25}{\micro\joule} output.} \\[8em]
\cite{Pierrot2013} & \color{red}SPMA/SPMA & \ce{Yb}/\SI{66}{\micro\m}/\SI{75}{\cm} & \SI{27}{\ps}/\SI{1031}{\nm} & \SI{780}{\fs}/\SI{63}{\MW} & \parbox{0.36\textwidth}{Parabolic pre-shaping. \SIadj{49}{\micro\joule} output.} \\[1em]
\cite{Fu2017} & \color{red}SPMA/SPMA & \ce{Yb}/\SI{34}{\micro\m} & \SI{9}{\ps}/\SI{1036}{\nm} & \SI{275}{\fs}/\SI{11}{\MW} & \parbox{0.36\textwidth}{Parabolic pre-shaping in a specialized 3C fiber. \SIadj{4.3}{\micro\joule} output.} \\[1em]
\cite{Wang2013} & SSA/GMNA & \ce{Yb}/\SI{11}{\micro\m}/\SI{2.2}{\m} & \SI{\sim1}{\ps}/\SI{1040}{\nm} & \SI{60}{\fs} & Optimized seed for SS. \\[1em]
\cite{Chen2012} & SSA/GMNA & \ce{Yb}/\SI{6.7}{\micro\m}/\SI{2}{\m} & \SI{200}{\fs}\textsuperscript{\footnotesize*}/\SI{1030}{\nm} & \SI{134}{\fs}/\SI{<16}{\kW}\textsuperscript{\footnotesize*} & \parbox{0.36\textwidth}{First investigation into pre-chirped amplification. \SIadj{2}{\nano\joule}\textsuperscript{\footnotesize*} output.} \\[1em]
\cite{Liu2015} & SPMA/GMNA & \ce{Yb}/\SI{90}{\micro\m}/\SI{1.2}{\m} & \SI{300}{\fs}/\SI{1030}{\nm} & \SI{60}{\fs}/\SI{2}{\MW} & \parbox{0.36\textwidth}{Pre-chirped amplification, which shows that both negative and positive chirps can be the optimal under different conditions. They pointed out that they were operating in the evolution before SSA regime, \ie in the SPMA regime. \SIadj{1.7}{\micro\joule} output.} \\[5em]
\cite{Song2017} & GMNA'/GMNA & \ce{Yb}/\SI{30}{\micro\m}/\SI{2.2}{\m} & \SI{>230}{\fs}/\SI{1040}{\nm} & \SI{24}{\fs}/\SI{31}{\MW}\textsuperscript{\footnotesize*} & \parbox{0.36\textwidth}{Pre-chirped amplification with counter-pumping. The \SIadj{24}{\fs} pulse has a small pedestal. \SIadj{1}{\micro\joule} output.} \\
\bottomrule
\multicolumn{6}{l}{\parbox{1.2\linewidth}{\vspace{0.3em}\footnotesize\textsuperscript{*} Not provided. Approximated from other values.\\ \hspace*{0.6em}Unprovided peak-power values are calculated by dividing pulse energy by pulse duration, regardless of Strehl ratio, so it is overrated.}} \\
\end{tabular}
\caption{Classification of prior works. Works that operate in their optimal nonlinear amplification regimes are colored in red. GMXA represents the intermediate regime between GMNA and GMSSA [Fig.~\ref{fig:regime_Yb_1050nm}(a)]. GMNA' or GMXA' with the ``prime'' symbol represents imperfect GMNA or GMXA, respectively.}
\label{tab:categorization1}
\end{table}

\clearpage 
\begin{table}[!ht]
\hspace{-10em}
\begin{tabular}{c|ccccl}
\toprule
\textbf{Prior work} & \parbox{0.18\textwidth}{\textbf{Its regime}/\\ \textbf{Optimal regime at this duration}} & \parbox{0.17\textwidth}{\textbf{Doped ion}/\\ \textbf{Core diameter}/\\ \textbf{Fiber length}} & \parbox{0.18\textwidth}{\textbf{Seed duration}/\\ \textbf{\hspace*{2em}wavelength}} & \parbox{0.267\textwidth}{\textbf{Compressed duration}/\\ \textbf{\hspace*{5em}peak power}} & \textbf{Notes} \\
\midrule
\cite{Zhao2014} & GMNA'/GMNA & \ce{Yb}/\SI{40}{\micro\m}/\SI{2}{\m} & \SI{360}{\fs}\textsuperscript{\footnotesize*}/\SI{1030}{\nm} & \SI{38}{\fs}/\SI{33}{\MW} & \parbox{0.36\textwidth}{Pre-chirped amplification with counter-pumping and a seed dechirped at \SI{190}{\fs}. \SIadj{1.3}{\micro\joule} output. It was previously categorized as SS.} \\[3em]
\cite{Liu2016b} & GMNA'/GMNA & \ce{Yb}/\SI{40}{\micro\m}/\SI{1.8}{\m} & \SI{>158}{\fs}/\SI{1038}{\nm} & \SI{33}{\fs}/\SI{51}{\MW} & \parbox{0.36\textwidth}{TOD-involved pre-chirped amplification with counter-pumping. \SIadj{1.7}{\micro\joule}\textsuperscript{\footnotesize*} output. It was previously categorized as SS.} \\[3em]
\cite{Luo2018} & GMXA'/GMXA & \ce{Yb}/\SI{40}{\micro\m}/\SI{2}{\m} & \SI{>180}{\fs}/\SI{1050}{\nm} & \SI{42}{\fs}/\SI{<10}{\MW}\textsuperscript{\footnotesize*} & \parbox{0.36\textwidth}{TOD-involved pre-chirped amplification with counter-pumping. \SIadj{440}{\nano\joule}\textsuperscript{\footnotesize*} output. It was previously categorized as SS.} \\[3em]
\cite{Wang2016c} & SSA/x & \ce{Yb}/\SI{29}{\micro\m}/\SI{2}{\m} & \SI{\sim120}{\fs}\textsuperscript{\footnotesize*}/\SI{1038}{\nm} & \SI{36}{\fs}/\SI{<11}{\MW}\textsuperscript{\footnotesize*} & \parbox{0.36\textwidth}{The broadband seed is beyond the scope of this study. Gain shaping is not clear due to significant output spectral energy around \SI{1030}{\nm}. \SIadj{390}{\nano\joule}\textsuperscript{\footnotesize*} output.} \\[3em]
\cite{Zhang2021b} & GMNA'/GMNA & \ce{Yb}/\SI{85}{\micro\m}/\SI{0.8}{\m} & \SI{200}{\fs}/\SI{1034}{\nm} & \SI{37}{\fs}/\SI{<55}{\MW}\textsuperscript{\footnotesize*} & \parbox{0.36\textwidth}{Double-pass pre-chirped amplification with counter-pumping. \SIadj{2}{\micro\joule}\textsuperscript{\footnotesize*} output.} \\[2em]
\cite{Zhang2020a} & GMNA'/GMNA & \ce{Yb}/\SI{85}{\micro\m}/\SI{0.8}{\m} & \SI{551}{\fs}/\SI{1036}{\nm} & \SI{47}{\fs}/\SI{<43}{\MW}\textsuperscript{\footnotesize*} & \parbox{0.36\textwidth}{Pre-chirped amplification with a circularly-polarized seed. \SIadj{2}{\micro\joule}\textsuperscript{\footnotesize*} output. It was previously categorized as SS.} \\[3em]
\cite{Li2024} & \parbox{0.1\textwidth}{{\color{red}SSA/SSA}\\(Fig.~\ref{fig:Nd_gain_spectrum})} & \ce{Nd}/\SI{4.5}{\micro\m}/\SI{3.5}{\m} & \SI{240}{\fs}\textsuperscript{\footnotesize*}/\SI{920}{\nm} & \SI{45}{\fs}/\SI{310}{\kW} & \parbox{0.36\textwidth}{Pre-chirped amplification with bidirectional-pumping and a seed dechirped at \SI{211}{\fs}. \SIadj{13}{\nano\joule} output.} \\[2em]
\cite{Wen2025} & SSA/GMSSA & \ce{Yb}/\SI{10}{\micro\m}/\SI{4}{\m} & \SI{>282}{\fs}/\SI{1096}{\nm} & \SI{47}{\fs}/\SI{480}{\kW} & \parbox{0.36\textwidth}{Pre-chirped amplification. The output pulse has low Strehl ratio (\num{\sim0.5}). This should be in the SSA regime, not SPMA, due to the generated broad spectrum around \SI{1100}{\nm} without deep spectral modulations typical of SPM. \SIadj{40}{\nano\joule} output.} \\
\bottomrule
\multicolumn{6}{l}{\parbox{1.2\linewidth}{\vspace{0.3em}\footnotesize\textsuperscript{*} Not provided. Approximated from other values.\\ \hspace*{0.6em}Unprovided peak-power values are calculated by dividing pulse energy by pulse duration, regardless of Strehl ratio, so it is over-estimated.}} \\
\end{tabular}
\caption{(Continued.) Classification of prior works. Works that operate in their optimal nonlinear amplification regimes are colored in red. GMXA represents the intermediate regime between GMNA and GMSSA [Fig.~\ref{fig:regime_Yb_1050nm}(a)]. GMNA' or GMXA' with the ``prime'' symbol represents imperfect GMNA or GMXA, respectively.}
\label{tab:categorization2}
\end{table}

\clearpage 
\begin{table}[!ht]
\hspace{-10em}
\begin{tabular}{c|ccccl}
\toprule
\textbf{Prior work} & \parbox{0.18\textwidth}{\textbf{Its regime}/\\ \textbf{Optimal regime at this duration}} & \parbox{0.17\textwidth}{\textbf{Doped ion}/\\ \textbf{Core diameter}/\\ \textbf{Fiber length}} & \parbox{0.18\textwidth}{\textbf{Seed duration}/\\ \textbf{\hspace*{2em}wavelength}} & \parbox{0.267\textwidth}{\textbf{Compressed duration}/\\ \textbf{\hspace*{5em}peak power}} & \textbf{Notes} \\
\midrule
\cite{Deng2009} & \color{red}GMNA/GMNA & \ce{Yb}/\SI{25}{\micro\m}/\SI{6}{\m} & \SI{200}{\fs}/\SI{1035}{\nm} & \SI{48}{\fs}/\SI{4.3}{\MW} & \parbox{0.36\textwidth}{Interestingly, the idea of this work realizing a broader gain for SS agrees quite well with GMNA, although it appeared ten years before the GMNA paper \cite{Sidorenko2019}. \SIadj{226}{\nano\joule} output.} \\[4em]
\cite{Sidorenko2019} & \color{red}GMNA/GMNA & \ce{Yb}/\SI{5}{\micro\m}/\SI{5}{\m} & \SI{700}{\fs}/\SI{1030}{\nm} & \SI{42}{\fs}/\SI{<1.8}{\MW}\textsuperscript{\footnotesize*} & \parbox{0.36\textwidth}{First investigation into \ce{Yb} GMNA. \SIadj{77}{\nano\joule} output.} \\[2em]
\cite{Krakowski2024} & \color{red}GMNA/GMNA & \ce{Er}/\SI{5.5}{\micro\m}/\SI{22}{\m} & \SI{1.21}{\ps}/\SI{1560}{\nm} & \SI{132}{\fs}/\SI{<45}{\kW}\textsuperscript{\footnotesize*} & \parbox{0.36\textwidth}{First investigation into \ce{Er} GMNA. \SIadj{6}{\nano\joule} output.} \\[2em]
\cite{Cooper2025} & \parbox{0.14\textwidth}{GMNA'/GMNA} & \ce{Yb}/\SI{50}{\micro\m}/\SI{1.5}{\m} & \SI{359}{\fs}/\SI{1030}{\m} & \SI{53}{\fs} & \parbox{0.36\textwidth}{Initial results aimed at scaling GMNA in a fiber with a \SIadj{50}{\micro\m} core. The fiber is not PM, which is expected to hinder the GMNA evolution \cite{Chen2023a}. Autocorrelation suggests a potentially low Strehl ratio, and the output spectrum deviates from the expected GMNA profile \cite{Sidorenko2019}. Nonetheless, this effort provides a valuable reference point for future investigations and is included here for completeness. \SIadj{2.9}{\micro\joule} output.} \\[8em]
\cite{Liu2017} & \color{red}GMNA/GMNA & \ce{Yb}/\SI{6}{\micro\m}/\SI{>2}{\m} & \SI{1030}{\nm} & \SI{40}{\fs}/\SI{1}{\MW} & \parbox{0.36\textwidth}{First high-peak-power (\ce{Yb}) Mamyshev oscillator, based on GMNA. It was previously categorized as SS. \SIadj{50}{\nano\joule} output.} \\[3em]
\cite{Olivier2019} & \color{red}GMNA/GMNA & \ce{Er}/\SI{5.5}{\micro\m}/\SI{10}{\m} & \SI{1555}{\nm} & \SI{93}{\fs}/\SI{126}{\kW} & \parbox{0.36\textwidth}{First \ce{Er} Mamyshev oscillator, based on GMNA. \SIadj{31}{\nano\joule} output.} \\[1em]
\cite{Boulanger2024} & \parbox{0.1\textwidth}{{\color{red}SSA/SSA}\\(Fig.~\ref{fig:Nd_gain_spectrum})} & \ce{Nd}/\SI{4.5}{\micro\m}/\SI{1.8}{\m} & \SI{920}{\nm} & \SI{53}{\fs}/\SI{122}{\kW} & \parbox{0.36\textwidth}{First \ce{Nd} Mamyshev oscillator. \SIadj{10}{\nano\joule} output.} \\
\bottomrule
\multicolumn{6}{l}{\parbox{1.2\linewidth}{\vspace{0.3em}\footnotesize\textsuperscript{*} Not provided. Approximated from other values.\\ \hspace*{0.6em}Unprovided peak-power values are calculated by dividing pulse energy by pulse duration, regardless of Strehl ratio, so it is overrated.}} \\
\end{tabular}
\caption{(Continued.) Classification of prior works. Works that operate in their optimal nonlinear amplification regimes are colored in red. GMXA represents the intermediate regime between GMNA and GMSSA [Fig.~\ref{fig:regime_Yb_1050nm}(a)]. GMNA' or GMXA' with the ``prime'' symbol represents imperfect GMNA or GMXA, respectively.}
\label{tab:categorization3}
\end{table}

\clearpage
\pagebreak
\bibliography{reference_supplement}